# A New Mathematical Framework for Atmospheric Blocking Events

Valerio Lucarini[1,2,3] [v.lucarini@reading.ac.uk], Andrey Gritsun[4]

1. Department of Mathematics and Statistics, University of Reading, Reading, UK
2. Centre for the Mathematics of Planet Earth, University of Reading, Reading, UK
3. CEN, University of Hamburg, Hamburg, Germany
4. Institute of Numerical Mathematics, Russian Academy of Sciences, Moscow, Russia

06/10/2019

## Abstract

We use a simple yet Earth-like hemispheric atmospheric model to propose a new framework for the mathematical properties of blocking events. Using finite-time Lyapunov exponents, we show that the occurrence of blockings is associated with conditions featuring anomalously high instability. Longer-lived blockings are very rare and have typically higher instability. In the case of Atlantic blockings, predictability is especially reduced at the onset and decay of the blocking event, while a relative increase of predictability is found in the mature phase. The opposite holds for Pacific blockings, for which predictability is lowest in the mature phase. Blockings are realised when the trajectory of the system is in the neighbourhood of a specific class of unstable periodic orbits (UPOs), natural modes of variability that cover the attractor the system. UPOs corresponding to blockings have, indeed, a higher degree of instability compared to UPOs associated with zonal flow.  Our results provide a rigorous justification for the classical Markov chains-based analysis of transitions between weather regimes. The analysis of UPOs elucidates that the model features a very severe violation of hyperbolicity, due to the presence of a substantial variability in the number of unstable dimensions, which explains why atmospheric states can differ a lot in term of their predictability. The resulting lack of robustness might be a fundamental cause contributing to the difficulty in representing blockings in numerical models and in predicting how their statistics will change as a result of climate change. This corresponds to fundamental issues limiting our ability to construct very accurate numerical models of the atmosphere, in term of predictability of the both the first and of the second kind in the sense of Lorenz.

# Table of Contents

# 1. Introduction

The dominant mechanism controlling the mid-latitude synoptic variability (time scales of 3-7 days) is the baroclinic instability, due to the presence of a strong equator-to-pole temperature difference. Baroclinic instability allows for the conversion of available potential energy into kinetic energy and generation of vorticity, which manifests itself into the phenomenology of mid-latitude cyclones (Charney 1947, Eady 1949). The regions of intense meridional temperature gradient coincide with the position of the jet stream aloft, where, in fact, the bulk of synoptic variability can be found. Baroclinically unstable modes transport heat poleward: this is one of the essential negative feedbacks determining the global stability of the atmospheric circulation. Baroclinic instability is also responsible for the error growth in weather forecast on synoptic spatial and temporal scales.

Low-frequency variability is associated with the vast range of atmospheric processes occurring on a time scale ranging from about a week to about a month, and , no complete understanding of its nature has yet been reached. At practical level, achieving efficient extended-range (beyond 7-10 days) forecast in the mid-latitudes is still very challenging and, on the climate side, understanding the impact of climate change on the low-frequency variability of the atmosphere is far from being fully understood; see, e.g., Ghil and Robertson (2002) and references therein.

Blockings are persistent, localized departures from the quasi-zonally symmetric flow in the mid-latitudes associated with the presence of large-amplitude, almost-stationary pressure anomalies (Rossby 1951), and are a key feature of the atmospheric low-frequency variability; see Tibaldi and Molteni (2018) and references therein. They are usually observed in either the Atlantic or in the Pacific sector, and, much more rarely, in both sectors at the same time (global blockings). The lifetime of a blocking event can range between few days and few weeks, and can lead to extreme and persistent anomalies in the local weather. Depending on the geographic location, on the season, on pre-existing conditions, phenomena as different as heat waves, cold spells, extreme dry conditions with extensive fires, and floods can occur as a result of blockings. Very relevant examples include the heat waves of 1976 in UK (Green, 1977) and of 2010 in Russia (Lau and Kim, 2011). The 2010 event, which led to a health crisis in Russia (Revich 2015), was dynamically linked to the occurrence of devastating rainfalls in Pakistan, thus indicating how the effects of blockings can cascade to regions far away from the actual high-pressure feature.

Blockings are *objectively* – so to say - identified through indices based on geopotential (Tibaldi and Molteni 1990), potential vorticity (PV) (Pelly and Hoskins 2003a) fields, statistical indicators based upon Empirical Orthogonal Functions (EOFs) (Barriopedro et al. 2006), and more sophisticated multidimensional approaches (Scherrer et al. 2005; Davini et al. 2012).

But, indeed, the phenomenology of blockings is very complex and will not be recapitulated here; see, e.g., Hoskins (1987) Pelly and Hoskins (2003a,b), Masato et al. (2012), Tibaldi and Molteni (2018), Woolings et al. (2018), and references therein.

## 1.1 Understanding Blocking Events

The mechanism(s) behind blocking events have been long investigated by the scientific community. We provide a brief summary below.

In the earlier literature, it was argued that the co-existence in the atmosphere of blocked and zonal states resulted from the presence of multiple stationary or quasi-stationary states, as codified in the classical theory of *Grosswetterlagen* (Namias 1968). Following the seminal paper by Charney and DeVore (1979), a great deal of interest was directed towards defining a minimal *theory of weather regimes* and finding confirmation of its validity by looking at observational data, see e.g. Legras and Ghil, 1985; Barnston and Livezey, 1987; Benzi et al. 1986; Ghil 1987; Ghil and Childress 1987, Mo and Ghil 1988, Benzi and Speranza 1989, Vautard 1990, Dymnikov and Kazantsev 1993; Vautard and Legras, 1998; Stan and Straus 2007; Ghil et al. 2018. Some investigations suggested that the processes behind low-frequency variability are of baroclinic, rather than barotropic, nature (Charney and Straus 1980, Cessi and Speranza 1985). The overall idea is that the blocked and zonal regimes correspond to fixed points, and the transitions between the two states result from the noise due to synoptic variability. Building on this, Dymnikov (1990) introduced dynamical indices, recently shown to predict well the blocking duration (Semenov et al. 2010). The theory of weather regimes is able explain quite accurately some nontrivial statistical properties of the mid-latitude atmosphere - see e.g. Ruti et al. (2006) - and can be used to look into the outputs of laboratory experiments. Using a rotating annulus, Weeks et al. (1997) and Tian et al. (2001) investigated the dynamical properties of a barotropic jet over topography and showed the existence of regimes resembling the zonal flow and the blockings observed in the mid-latitude atmosphere. They discovered that in a range of values of the Rossby number, the two regimes coexist, with the barotropic flow jumping erratically from the one to the other, as in the barotropic model on the sphere of Legras and Ghil (1985).

Blockings have also been interpreted as isolated, coherent structures, corresponding to special weakly (Malguzzi and Malanotte-Rizzoli, 1984) or fully nonlinear (Flierl, 1980; McWilliams, 1980, Haines and Marshall, 1987) stationary solutions - modons or vortex pairs - of the inviscid and unforced quasi-geostrophic (QG) equations on the rotating sphere; see Butchart et al. (1989) for a summary of this point of view.

Following statistical mechanics and dynamical systems theory, Ghil (1987), Mo and Ghil (1988), Vautard et al. (1990) and Kimoto and Ghil (1993a,b) proposed to model the time evolution of the coarse-grained state of the atmosphere using Markov chains, each weather regime being identified with one of the states of the chain. As a result, the Markov transition matrix defines the probability of transition from one regime to another one. Subsequently, Deloncle et al. (2007) and Kondrashov et al. (2007) proposed the use of random forests algorithms to find the best predictors for the transitions. The Markov chain approach allows for studying simultaneously the statistics of each regime and the paths of transitions among them; see also Franzke et al. (2008, 2011) and Tantet et al. (2015).

A rather different point of view, building upon the differences between the Atlantic and Pacific blockings, focuses on understanding the local causes for blockings, interpreted as forced structures resulting from eddy forcings (Haines and Marshall, 1987) or from Rossby wave breaking (Pelly and Hoskins 2003a), rather than actual stationary states. Additionally, it was emphasized the importance of looking at the relationship between blocking events and teleconnection patterns. Croci-Maspoli et al. (2007) and Athanasiadis et al. (2010) showed that there is correlation between blocking occurrence and the phase of the NAO, the North Atlantic Oscillation (Pacific North American teleconnection pattern, PNA) in the Atlantic (Pacific) sector. Nakamura et al. (1997) argued that while Atlantic blockings are associated with forcings taking place in the region of low frequency variability, the Pacific ones are more directly related to high frequency, synoptic activity. It is also known that the statistics of blockings, especially in the Atlantic sector, features a rather high sensitivity to forcings, in the form of a substantial interannual and multidecadal variability associated with relatively weak anomalies in the solar forcing (Rimbu and Lohmann 2011; Rimbu et al. 2014) or in the heat exchange between atmosphere and the ocean (Häkkinen et al. 2011).

Numerical weather prediction systems are routinely benchmarked against their ability to predict onset and decay and statistics of blocking events (Ferranti et al. 2015). Note that it is usually assumed that the predictability of weather during blocking events is higher than average, while the extremely challenging task is, specifically, to predict the onset (Mauritsen et al. 2004) and decay (Quandt et al. 2017) of blockings. See also discussion in Oortwijn (1998) and in Pelly and Hoskins (2003b).

The complexity of blockings is emphasized by the fact that current climate models show a (relatively) limited skill in simulating them (Scaife et al. 2010, Barriopedro et al. 2010), marginally improving in the last two decades (Davini and D'Andrea, 2016). In turn, atmospheric blockings could play an important role in defining the future climate: Masato et al. (2013) suggested that the 2010 Russian blocking could, in future, become a dominant regime. A foreseen impact of climate change is the reduction on the equator-to-pole temperature gradient, which has been interpreted as possibly leading increased blocking frequency (Francis et al. 2012), in agreement with classical arguments by, e.g., Charney and DeVore, 1979; Ghil and Childress, 1987; Legras and Ghil, 1985). Contrasting points of view have been presented in the literature; see, e.g., Hassanzadeh et al. (2014). Woolings et al. (2018) summarize the state of the art of possible impacts of climate change on the properties of blockings events.

### 1.2 A different mathematical framework
Some authors suggested that the classic identification of recurrent weather regimes with fixed points was conceptually unsatisfactory, and indicated that one should try instead to take into direct account the chaotic nature of the atmosphere (Speranza and Malguzzi 1988; Malguzzi et al. 1990). In a recent numerical investigation performed using a QG model, Schubert and Lucarini (2016), used the formalism of Lyapunov exponents (LEs) and covariant Lyapunov vectors (CLVs) (Eckmann and Ruelle 1986; Ginelli et al. 2007; Wolfe and Samuelson 2007) to study the linear stability properties of the turbulent background flow. They found that when blocking occurs, the global growth rates of the fastest growing CLVs are significantly higher. Hence, against intuition, the instability is stronger during the blocked phases. Such an increase in the finite time LEs (FTLEs) with respect to typical, zonal conditions is attributed to a combination of stronger barotropic and baroclinic conversions, see also an earlier analysis by Frederiksen and Bell (1990). Schubert and Lucarini (2016) interpreted such a counter-intuitive finding – blockings are usually thought as being characterized by anomalously high predictability  - as resulting from the difficulty of predicting the specific timing of onset (Mauritsen et al. 2004) and decay (Quandt et al. 2017) of the blocking event. Kwasniok (2018, personal communication) found that atmospheric flows associated with anomalously high values of finite time largest LE resemble correspond to blocked conditions. Vannitsem (2001), showed that the atmospheric patterns characterised by high instability are closely related to the negative phase of the PNA. Considering the link discussed above between PNA and blocking statistics, this finding is also in agreement with Schubert and Lucarini (2016). Agreement is also found with what has been recently proposed by Faranda et al. (2016, 2017) who, using methods borrowed from extreme value theory for dynamical systems (Lucarini et al. 2016) identified blocking regimes with unstable fixed points in a severely projected phase space, and derived that blockings come higher instability of the circulation, associated with higher effective dimensionality of the system.

The higher instability of the blocked vs. the zonal state agrees with the findings of Weeks et al. (1997) and Tian et al. (2001), who discovered that, in the parametric regime where zonal and blocked states coexist for the barotropic jet over topography, the blocked states featured a much larger variability than the zonal ones.

Here, we will take the statistical mechanical point of view that sees climate as a non-equilibrium steady state system (NESS), see Lucarini et al. (2014, 2017). Instead of trying to construct a heavily truncated, low-dimensional atmospheric model and look at the (possibly stochastically perturbed) stationary solutions we will extract from the complex high-dimensional dynamics of a model its essential building blocks, true nonlinear modes. Such building blocks are the so-called unstable periodic orbits (UPOs) of the system (Cvitanovic 1988, 1991), which populate the attractor of a chaotic system. They can be seen as the generalisation of the normal modes one finds in, e.g., a network of coupled linear oscillators

UPOs define the so-called skeletal dynamics and, since they populate densely the attractor, can be used to reconstruct all of the statistical properties of the system (Grebogi et al. 1988). While constructing unstable closed orbits in a high-dimensional system seems an unfeasible task, UPOs are widely used to study complex systems (Cvitanovic 2013; Cvitanovic et al. 2016). Following the early work by Kawahara and Kida (2001) who found an UPO in a Navier-Stoke simulation of a plane Couette flow and showed that just one UPO was able to describe in a surprising accurate way the statistical properties of the turbulent flow, UPOs-based approach have shown a great potential for proving an alternative approach for the study of the properties of turbulence, see e.g. Kreilos and Eckhart 2012, Willis et al. 2013). UPOs have also shown their potential for studying simple barotropic flow featuring low-dimensional (Kazantsev 1998) and high-dimensional (Gritsun 2008; 2013) chaos, and for understanding non-trivial resonant responses ("climate surprises") of the model to forcings showing a violation of the fluctuation-dissipation theorem (Gritsun and Lucarini 2017).

## 1.3 This paper

In this paper we wish to advance the mathematical understanding of blockings and reconcile some of the dynamical points of view proposed so far, in order to clarify their properties in terms of predictability, to understand to what extent blockings can be associated with specific modes of the atmospheric circulation (and define such modes), and in order to shed light on the fundamental reasons why it is so hard to construct numerical models able to represent them correctly. This will require abandoning the paradigm of low-dimensional dynamical systems, take into account the need to look beyond stationary solutions, and adopt the viewpoint of non-equilibrium statistical mechanics.

We will perform our analysis on the classical Marshall and Molteni (1993) model, which provides a basic yet solid framework for understanding the synoptic-to-planetary scale dynamics of the atmosphere and has been specifically designed for studying its low-frequency variability, and it is quite successful at this regard (Corti et al. 1997, Michelangeli and Vautard 1998, Vannitsem 2001). While such a model is far from being – in any sense – realistic, we take it as a very meaningful starting point for our analysis of the mathematics of the atmosphere.

We will look, on the one side, at the properties of the tangent space of the system using the formalism of LEs and CLVs, and, on the other side, reconstruct the skeletal dynamics by computing UPOs. The idea we propose is to create a high-dimensional counterpart of the classic theory of weather regimes, associate specific UPOs to Atlantic, Pacific, and global blockings. We remark another important aspect we will explore is connected to structural stability properties. When a system has UPOs with different number of unstable directions, the number of dimensions associated with locally growing modes is not constant on the attractor; specifically, the system is not hyperbolic, hence not structurally stable (Lai et al. 1997, Kostelich et al. 1997). This implies that small perturbations to the dynamics can lead to substantial changes in the statistical properties of the system. Additional problems emerge because for a system possessing variability in the number of unstable dimensions, numerical implementations provides an output that is unlikely to be close to the true trajectories (shadowing property) for arbitrarily long times (Do and Lai, 2004), as opposed to the hyperbolic dynamical systems (Katok and Hassenblatt, 2003). We will show that blockings play a prominent role in determining the heterogeneity of the attractor. This could be a major issue underpinning the above-mentioned difficulties in representing blockings in numerical weather prediction and climate models and in making climate change projections for the statistics of blockings.

The paper is structured as follows. In Section 2 we present our model, describe its structure and its main properties, and summarize its performance in terms of representation of the Atlantic, Pacific, and Global blocking events. In Section 3 we show how we can reinterpret blocking events and clarify their mathematical nature using LEs and CLVs to study their predictability, and link them to rigorously defined atmospheric modes, i.e. recurrent weather patterns defined by UPOs. We will see whether there is something *really special* about blocking events. We will investigate the extent of the variability of the unstable dimensions in the atmosphere, and how this affects blocking events. In Section 4 we provide our conclusions and perspectives for future works. In the Appendix we provide a somewhat informal introduction to all the mathematical concepts needed to follow the presentation and discussion of our results. The reader who has solid knowledge of dynamical systems theory can skim through the Appendix.

## 2. The Marshall-Molteni Model

We perform our simulations using the popular Marshall-Molteni (1993) model, which provides a parsimonious but quite effective representation of the synoptic-to-planetary scale atmospheric dynamics of the mid-latitudes. This model is constructed by taking the QG approximation for atmospheric dynamics (Holton and Hakim 2013) and considering a coarse vertical discretisation of the atmosphere in three layers centered at the 200, 500 and 800 hPa pressure levels (layers 1, 2, and 3, respectively). The dynamics of each layer $j$ is described by the evolution equation of the QG potential vorticity $q_j$, j=1,2,3 defined as follows:

$$q_1 = \Delta\psi_1 - (\psi_1 - \psi_2)/R_1^2 + f \qquad (1a)$$
$$q_2 = \Delta\psi_2 + (\psi_1 - \psi_2)/R_1^2 - (\psi_2 - \psi_3)/R_2^2 + f \qquad (1b)$$
$$q_3 = \Delta\psi_3 + (\psi_2 - \psi_3)/R_2^2 + f(1 + h/H_0) \qquad (1c)$$

where $\Delta$ is the horizontal Laplacian operator, $\psi_j$ is the streamfunction at the level j (such that $\vec{u}_j = \vec{\nabla}^\perp \psi_j$ is the geostrophic wind at the level j), $R_j$ is the Rossby deformation radius, defining the vertical dynamical coupling between the level j and j+1, $f$ is the latitude dependent Coriolis constant, $h$ is the orography of the surface, rescaled by the constant $H_0$. We remind that, given the nature of the geostrophic approximation, the temperature field $T_k$ is defined at the levels k=1.5 and k=2.5 (i.e. intermediate between those defining the levels of pressure), with $T_k \propto \psi_{k+\frac{1}{2}} - \psi_{k-\frac{1}{2}}$. The equations describing the evolution of the model are defined as follows:

$$\partial_t q_j + J(\psi_j, q_j) = -D_j + S_j, \qquad j = 1,2,3 \qquad (2)$$

where $J(A, B)$ is the Jacobian operator defining the nonlinear advection. At each level $j$, instead, $-D_j$ is the operator defining the linear damping and friction acting and leading to a contraction of the phase space volume and $S_j$ is the forcing term, injecting energy into the system. The dissipative terms are defined as follows:

$$-D_1 = (\psi_1 - \psi_2)/(\tau_R R_1^2) - R^8 \Delta^4 \acute{q}_1/(\tau_H \lambda_{max}^4) \qquad (3a)$$
$$-D_2 = -(\psi_1 - \psi_2)/(\tau_R R_1^2) + (\psi_2 - \psi_3)/(\tau_R R_2^2) - R^8 \Delta^4 \acute{q}_2/(\tau_H \lambda_{max}^4) \qquad (3b)$$
$$-D_3 = -(\psi_2 - \psi_3)/(\tau_R R_2^2) - EK_3 - R^8 \Delta^4 \acute{q}_3/(\tau_H \lambda_{max}^4) \qquad (3c)$$

where $\tau_R$ is the radiative relaxation timescale; $\tau_H$ is the timescale of hyper diffusion; $R$ is the Earth's radius; and $\lambda_{max}$ is the absolute value of the largest eigenvalue of the Laplacian in the model grid ($\lambda_{max} = 18\times19$ for T18, see below). In particular, $EK_3 = \nabla \cdot (k_{surf}\nabla\psi_3)$ is the surface friction, with $k_{surf} = (1 + 0.5LS + 0.5FH(h))/\tau_E$, where LS is the fraction of land in the surface gridbox, $FH(h) = 1 - \exp(-h/1000m)$, and $\tau_E$ is the Ekman friction timescale. Finally, $\acute{q}_i = q_i - f, i = 1,2$, and $\acute{q}_3 = q_3 - f(1 + h/H_0)$.

Once we have defined the functional forms of $D_j$, one can engineer the source terms $S_j$ that give the system – to a very first approximation – a dynamical behaviour similar to what observed in nature. The idea is to take the long term average of Eq. (2):

$$S_j = \overline{J(\psi_J, q_J)} + \overline{D}_J, \qquad (4)$$

and insert on the right hand side actual atmospheric data in the form of streamfunction and QG potential vorticity at the desired atmospheric levels.

Table 1: Summary of the main features of the model used here vs. the original version (Marshall and Molteni 1993).

|  | Marshall and Molteni (1993) | This paper |
|---|---|---|
| Area | Global | Northern Hemisphere |
| Truncation | T21 | T18 |
| Number of degrees of freedom | 1449 | 513 |
| Data for RHS | ECMWF analyses, JF 1984-1992 | ERA40, DJF 1983-92 |
| $H_0$ | 9 km | 8 km |
| $\tau_E$ | 3 day | 1.5 day |
| $\tau_H$ | 2 day at m=21 | 1.33 day at m=18 |
| $\tau_R$ | 25 day | 30 day |
| time step | 1/40 day | 1/40 day |
| $R_1$ | 700 km | 761 km |
| $R_2$ | 450 km | 488 km |
| Number of positive LEs | 154 | 37 |
| $\lambda_1$ | ≈0.24 day$^{-1}$ | ≈0.14 day$^{-1}$ |
| $d_{KY}$ | ≈389 | ≈89.1 |
| $h_{KS}$ | ≈11.2 day$^{-1}$ | ≈2.13 day$^{-1}$ |

The resulting time-independent fields $S_j$ can then be then used as forcing acting in the model. $S_j$ constrains the solution of the model to be statistically stable and close to the climatology of the atmospheric fields used as input. The data we use to construct $S_j$ are drawn from the 1983-1992 winter (DJF) climatology of the ERA40 reanalysis provided by ECMWF (Uppala et al. 2005). This provides a rough yet effective way to force our simple atmospheric system to resemble actual winter atmospheric conditions.

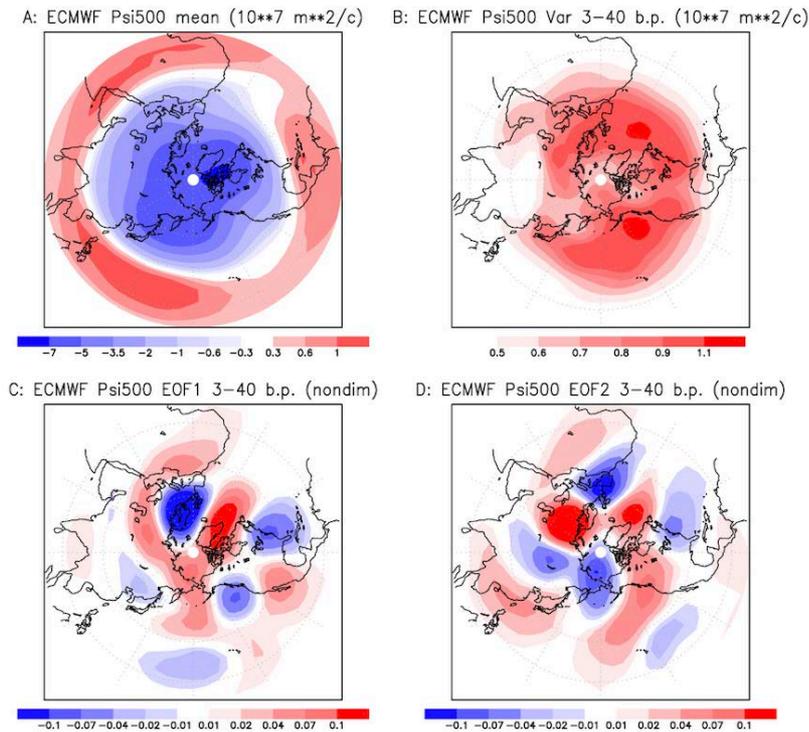

**Figure 1: Statistical properties of the ECMWF reconstruction of the atmosphere. We focus on the northern hemisphere and the 500 hPa level. a) Mean and b) Variance of the Streamfunction in the frequency band (3day)$^{-1}$-(40days)$^{-1}$. c) First and d) Second EOF in the frequency band (3day)$^{-1}$-(40days)$^{-1}$.**

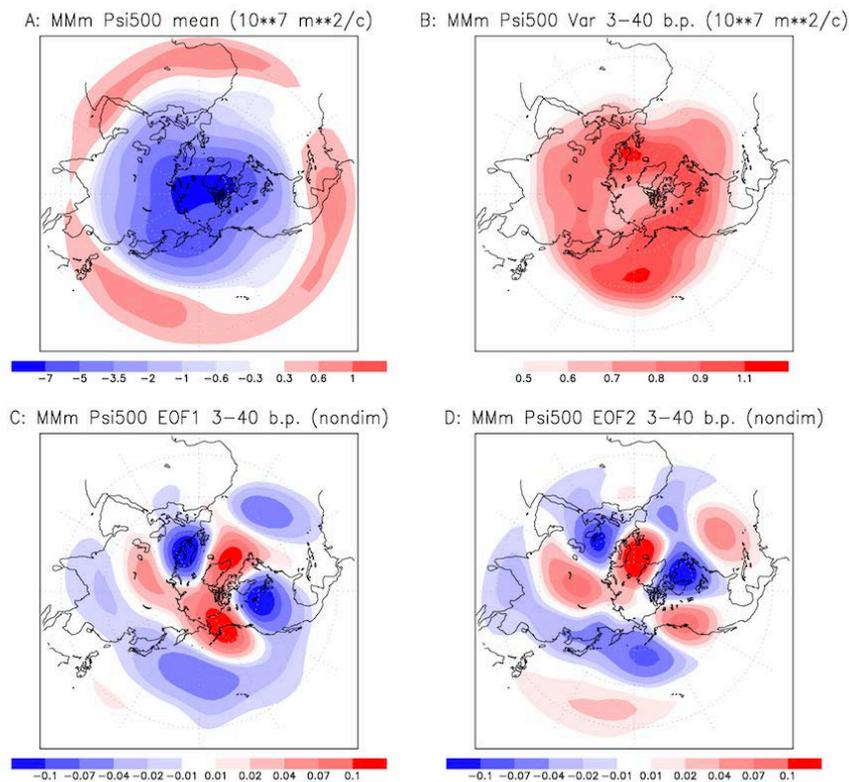

**Figure 2: Same as Fig. 1, for the Marshall-Molteni model used in this investigation. Variance (panel b) is here multiplied by a factor 1.25. The model is forced using the ECMWF data; see text for details.**

We run the model adopting a T18 horizontal resolution (corresponding to 54 lon x 27 lat gridpoints globally) for a total of 125000 days after discarding an initial transient, and further restrict the model's domain to the Northern Hemisphere. The choice of a low resolution (compared to T21 usually used in many studies performed with this model) and of a limited domain is motivated by the need to study accurately the tangent space, and, especially, by our desire to compute UPOs. Finding and estimating accurately UPOs has a computational cost that increases exponentially with the dimensionality of the system, see the Appendix. Choosing parameters for the system that are conducive to having stronger dissipation, thus leading to a weaker instabilities and lower Kaplan-Yorke (KY) dimension for the attractor contributes to making the job of finding UPOs somewhat easier. In Table 1 we summarise the main features of the model used in this paper.

Despite the strong simplifications associated with the low resolution and the use of QG approximation, the overall skill of the Marshall-Molteni model in representing the synoptic and planetary scale atmospheric variability is rather good (Corti et al. 1997, Michelangeli and Vautard 1998, Vannitsem and Nicolis 1997, Vannitsem 2001). Despite the further simplifications adopted here, the model does a fairly good job in representing the main features of the Northern Hemisphere atmosphere. Figure 1 portrays the mean (Panel a) and variance (Panel b) of the streamfunction, as well as the first and second EOFs at 500 hPa for the ERA40 fields used to construct the forcing for the model, while Fig. 2 shows, correspondingly, the output of the model. Even if the forcing is constructed using mean fields, the structure of variability of the true atmospheric system is captured fairly well. Model has somewhat reduced (by about 20%) level of variability, see caption of Fig. 2. We can find a strong signature of PNA (negative phase) in the Pacific sector of EOF1.

## 2.1 Tibaldi-Molteni Index for Blocking Events

We describe briefly the adapted Tibaldi and Molteni (1990) for detecting blockings in our model. We study the occurrence of reversals in the direction of the zonal wind at a given longitude $\lambda$ with respect to normal conditions in the mid-latitudes latitudinal band $[\phi_S, \phi_N]$ centered on $\phi_0$. We choose $\phi_S = 40^o N$, $\phi_0 = 60^o N$, and $\phi_N = 80^o N$. We first construct the geopotential field $Z_j = f_0/g\ \psi_j$ where $g$ is the gravity acceleration and $f_0 = \sqrt{3}\Omega$ is the reference Coriolis parameter at $60^o N$, with $\Omega = 2\pi/day$. Then, we look at the level $j = 2$ (500 hPa), and construct for each longitude $\lambda$ the following time series:

$$GHGN(\lambda, t) = \frac{Z_2(\phi_N + \delta\phi, \lambda, t) - Z_2(\phi_0 + \delta\phi, \lambda, t)}{\phi_N - \phi_0} \qquad (5a)$$

$$GHGS(\lambda, t) = \frac{Z_2(\phi_0 + \delta\phi, \lambda, t) - Z_2(\phi_S + \delta\phi, \lambda, t)}{\phi_0 - \phi_S} \qquad (5b)$$

We consider a particular longitude $\lambda$ blocked at time $t$ if $GHGS(\lambda, t) > 0$ and $GHGN(\lambda, t) < -12 m/\ ^o lat$ for at least one value of $\delta\phi = -4^o, 0, 4^o$. Note that we our criterion is slightly more stringent than in Tibaldi and Molteni (1990), because we want to focus on stronger blockings. It is commonly assumed that we can identify a true blocking if these conditions persist for at least two days. We say that we are experiencing an Atlantic (a Pacific) blocking when at least one $\lambda \in [56^oW, 80^oE]$, ($\lambda \in [104^oE, 90^oW]$) has, while we have a global blocking when we the Atlantic and the Pacific are simultaneously blocked. Note that blocked conditions usually have long spatial correlations, i.e. they extend over many degree of longitude, as they correspond to large-scale, quasi-stationary atmospheric patterns.

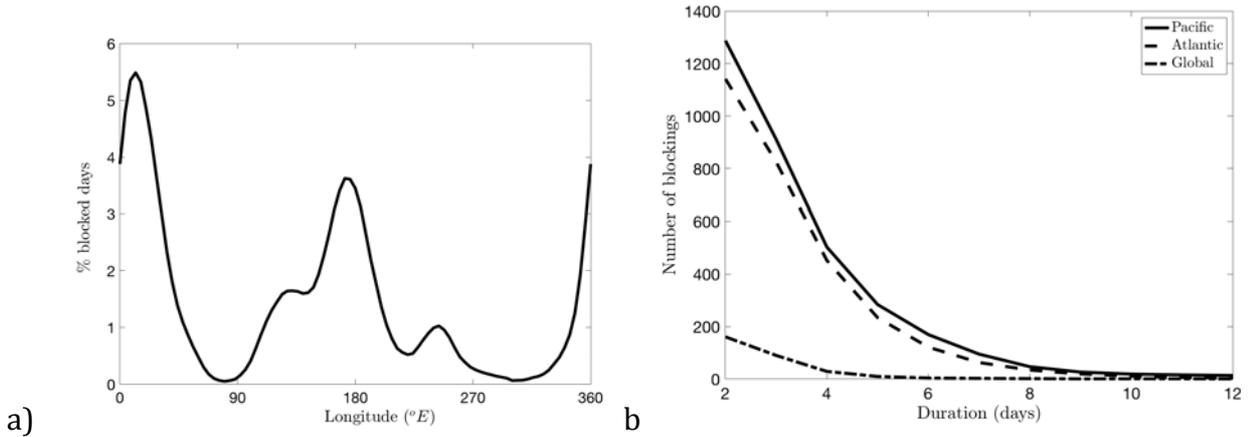

a) b)

**Figure 3:** a) Geographical prevalence of blocked states. b) Statistics of the blocking events as a function of their duration (in days). Solid line: Pacific blockings. Dashed line: Atlantic blockings. Dash-dotted line: Global blockings.

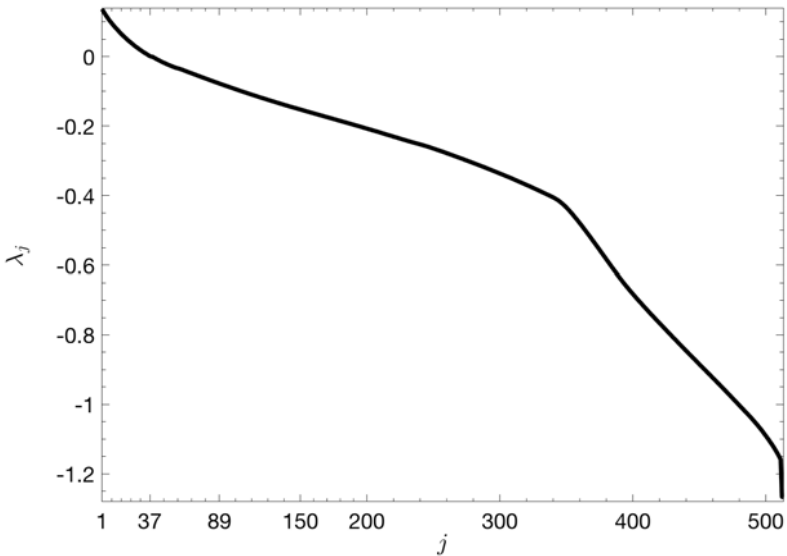

**Figure 4:** Spectrum of Lyapunov exponents for the model in the configuration described in Table 1. The number of positive exponent (37) and the integer part of the KY dimension (89) are indicated.

## 3. Results

The Marshall-Molteni model provides a good representation of the geographical prevalence of blocking events. Figure 3a shows that blockings occur mostly in the Atlantic and Pacific sectors. We have found a total of 6550 blocking events in 125000 days of simulation, of which about 3350 are in the Pacific sector, 2900 are in the Atlantic sector, and 300 are global. Looking at Fig. 3a, we find that the we have a clear quantitative underestimate with respect to observations by a factor of about 3 (Tibaldi and Molteni 2018); note also that, as mentioned above, we use here more stringent conditions than usually done for defining the occurrence of a blocking.

Figure 3b) shows the distribution of blocking events according to their lifetime. The distribution is approximately exponential (thus not featuring the more complex structure described by Pelly and Hoskins 2003a), and only few blockings have a lifetime longer than 7 days. Note that using the standard T21 configuration for the Marshall and Molteni model one obtains a fraction of blocked dates larger by a factor of about 1.5 for the Atlantic and Pacific sector and by a factor of about 4 for the global blockings. Nonetheless, the statistics with respect to time duration and geographical prevalence is similar to what obtained with the T18 model (not shown).

### 3.1 Dynamical Heterogeneity of the Attractor

Figure 4 portrays the spectrum of LEs for the model described above, using the parameters given in Table 1. The model features high-dimensional chaos, as we find 37 positive LEs and a KY dimension of about 89.1. We remark that very large in absolute value negative LEs found for indices larger than about 320 are associated with the hyperdiffusion operator included in the equations of motion. We remark that Vannitsem and Nicolis (1997) provided the first analysis of the instabilities of the Marshall and Molteni (1993) model using the formalism of LEs.

The LEs provide extremely useful yet limited information on the instability of the atmosphere, because they describe asymptotic, averaged rates of exponential increase or decrease of infinitesimal perturbations. We analyse the heterogeneity of the attractor by evaluating the variability of the properties of the tangent space as captured along a trajectory lasting 125000 days. We remark that we compute all the dynamical indicators of the tangent space using the minimal time scale allowed by our model, i.e. its time step of 1/40 day, but we report daily averages. All the mathematical concepts and terminology used in the following and details on the actual computations are given in the Appendix.

In Fig. 5a, in agreement with previous investigations performed using both more complex (De Cruz et al. 2018) and simpler (Vannitsem and Lucarini 2016) models, we find that the daily value of the first FTLE fluctuates substantially along the trajectory, with mean value corresponding to the asymptotic value for the first LE given in Table 1. Note that, by construction, the first FTLEs is the same whether we use the covariant or the backward definition. We find non-negligible presence of negative values, which could be due to the violation of hyperbolicity in the system. Nonetheless, the system does not have regions where we have return of skill: if, instead we plot the daily averages of the largest covariant or backward FTLE we find a broad distribution with exclusively positive support (not shown). Note that the problem of assessing whether a system is uniformly hyperbolic is usually addressed by studying whether the unstable and stable tangent spaces have tangencies (see e.g.. Ginelli et al. 2007; Kuptsov and Kuznetsov 2018). We will address this matter using the formalism of UPOs, see Sect. 3.3.

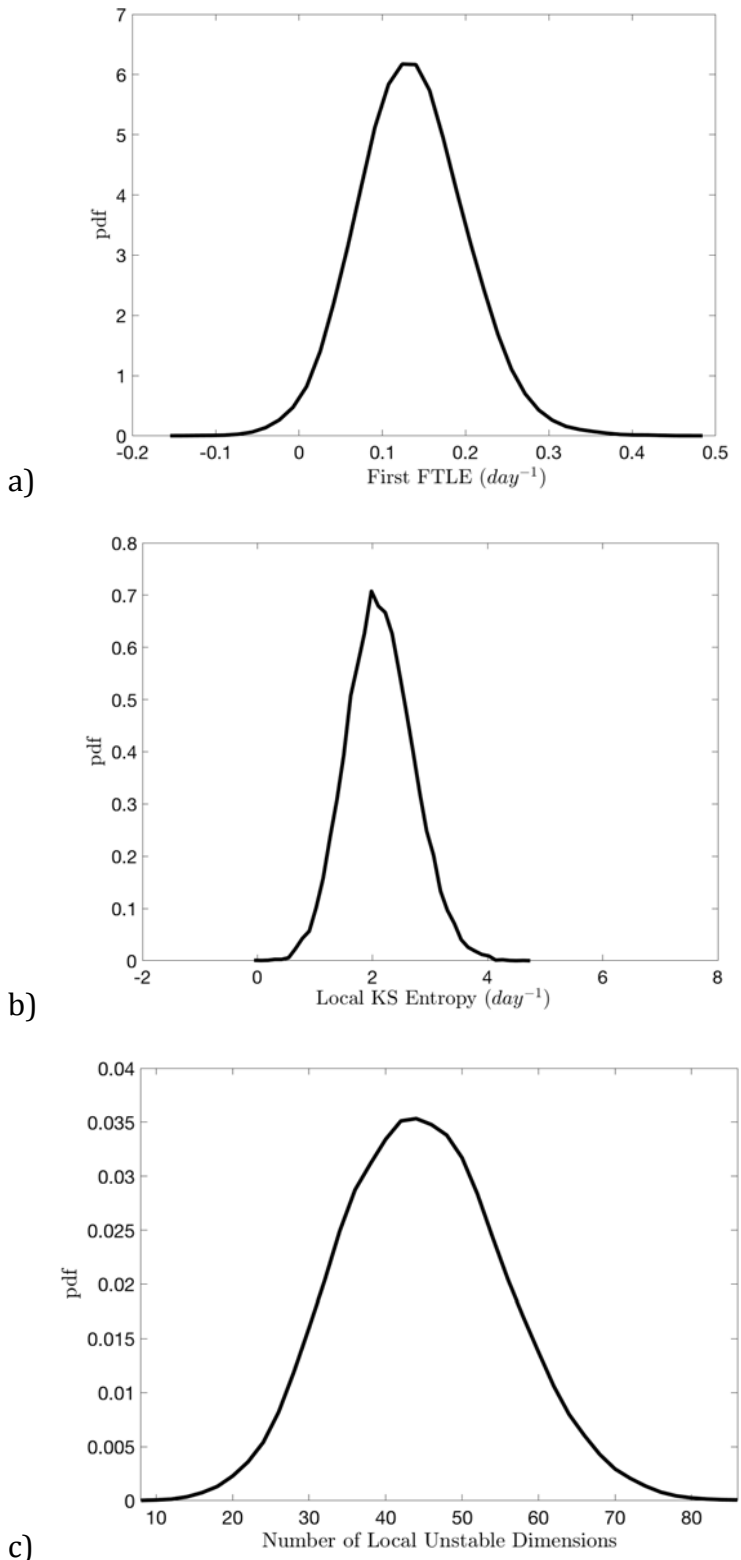

a)

b)

c)

Figure 5: Heterogeneity of the attractor of the model. Pdf's of instantaneous value of dynamical indicators describing the instability of the system. a): First finite-time Lyapunov exponent. b) Local estimate of the Kolmogorov-Sinai entropy.. c) Number of local unstable dimensions.

We can better appreciate how heterogeneous the tangent space is by noting – see Fig. 5b – that the distribution of the local estimate of the Kolmogorov-Sinai (KS) entropy, given by the daily average of the sum of the first 37 backward FTLEs corresponding to the unstable tangent space, does not have positive support. In other terms, there are instances where several asymptotically unstable modes become locally stable. While the mean value of the distribution corresponds to the asymptotic value of $h_{KS}$ reported in Table 1, the local estimate of the KS entropy features very large fluctuations, with the low-end tail, which includes negative values, accompanied by positive anomalies corresponding to strongly enhanced instability. Such variability results from the fact that several FTLEs fluctuate between positive and negative values, and from the fact that such fluctuations are partially coherent across the spectrum of exponents. This can be appreciated by looking at Fig. 5c, where we show that the number of unstable dimensions fluctuates by almost an order of magnitude. Note that, for the reasons explained in the Appendix on how averages are computed, the mean value of the dynamical indicator shown in Fig. 5c is somewhat larger than (yet broadly in agreement with) the asymptotic values reported in Table 1.

**3.2 Instability of the Atmosphere during Blocking Events**
We now wish to test in detail the idea proposed by Schubert and Lucarini (2016) that blockings are, on the average, associated to conditions of anomalously high instability as a result of the difficulty in predicting their onset and decay. We then compute the dynamical indicators of instability during each blocking event; we stratify the results by aggregating the statistics of blocking events of the same temporal length, as defined using the Tibaldi-Molteni index. The analysis is performed separately for Atlantic, Pacific, and global blockings, and results are shown in Figs. 6-9.

In Fig. 6a-c we look into the life cycle of Atlantic blocking events. We use a relative time axis where day 5 corresponds to the onset of the blocking, and day 5+d corresponds to its decay. We shows results for duration of d= 3 days (about 850 events), d = 5 days (about 230 events) and d= 6 to 8 days (about 210 events), and d=9 to 12 days (about 35 events). We include error bars corresponding to the $\pm 1\sigma$ confidence interval for the average value of the dynamical indicators for blocking of specified duration; this means that we are dividing the standard deviation of the samples by the square root of the number of blockings.

We find that blocking events are indeed associated with regions in the phase space of the systems where the dimensionality of the unstable manifold is higher than average (Fig. 6c). While the average number of local unstable dimensions is about 46, during blocking events the number grows to about 48-50. We find evidence of the fact that the positive anomalies in the number of unstable dimension is highest at the onset and decay phase of the blocking, while in the mature stage we have a relative minimum, with a smaller (yet positive) deviation with respect to the long term average. The fact that during the mature phase, the instability is reduced provides a good match with the standard interpretation of blocking as a phase of enhanced predictability of the weather. This interpretation is confirmed when looking at panel b, where we show how the local estimate of the KS entropy changes during the life cycle of the blocking events. We find that, on one side, instability is in general higher during blocking events than in typical conditions. On the other side, during the blocking events, instability is largest at the onset and decay phases. Finally, Fig. 6a shows the daily values of the first FTLE, which is associated with the direction featuring the fastest error growth on a 1-day time scale. While this dynamical indicator is considerably easier to compute that the other ones presented before, its time evolution during the life cycle of the blocking event is in somewhat worse agreement with what discussed earlier. This indeed suggests that blocking events are associated with complex, multiscale dynamical instabilities.

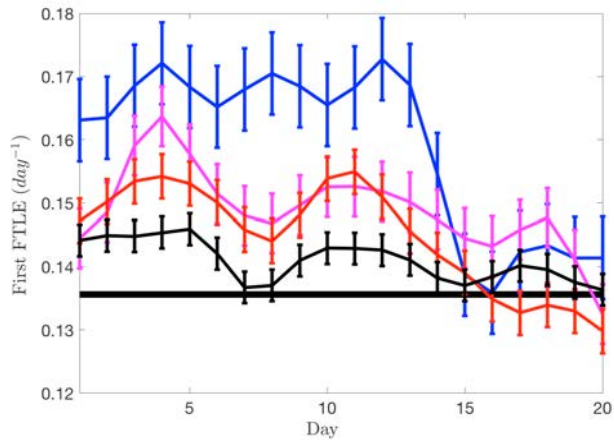

a)

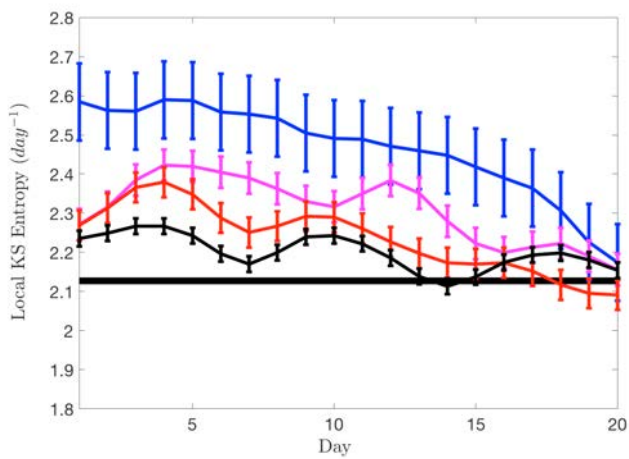

b)

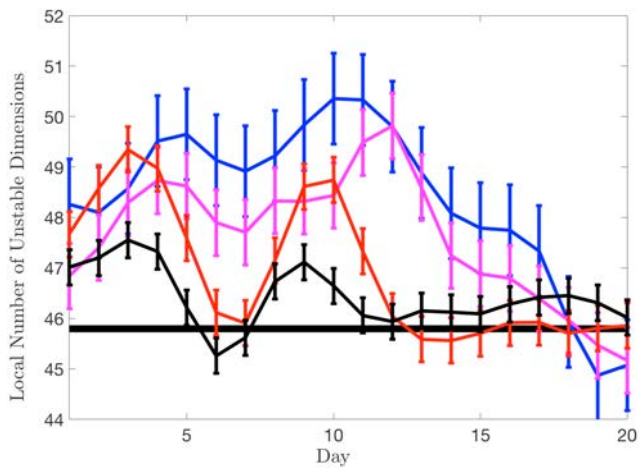

c)

**Figure 6: Life cycle of Atlantic blockings. Onset takes place at day 5. Thick horizontal solid line: ensemble average. Black line with error bars: 3-day blocking events. Red line with error bars: 5-day blocking events. Magenta line with error bars: 6 to 8-day blocking events. Blue line with error bars: 8 to12-day blocking events. a) First finite-time Lyapunov exponent. b): Local estimates of the Kolmogorov-Sinai entropy. c) Number of local unstable dimensions. See text for details.**

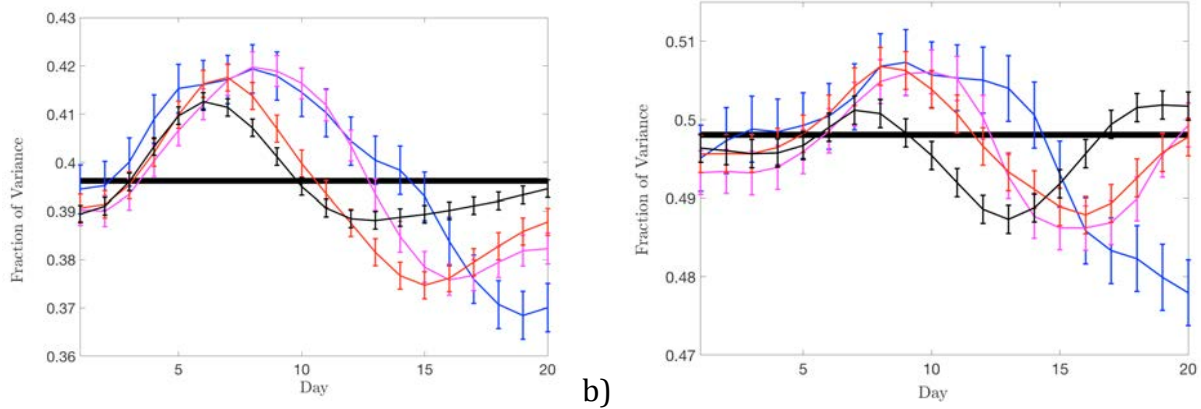

**Figure 7: Panel a) Average projection of the five leading unstable CLVs on the Atlantic Sector during the life cycle of Atlantic blockings. Onset takes place at day 5. Thick horizontal solid line: ensemble average. Black line with error bars: 3-day blocking events. Red line with error bars: 5-day blocking events. Magenta line with error bars: 6 to 8-day blocking events. Blue line with error bars: 8 to12-day blocking events. B) Same as a), but for Pacific blockings.**

In Fig. 7a we show how the average projection on the Atlantic sector (more precisely, the geographically-restricted L2 norm) of the first five CLVs changes during the life cycle of Atlantic blocking events. We find that such projection is higher than average during the blocking, and is lower than average just before and after the event. Note also that the time evolution of the projections considered here flag the life cycle of the blocking events in good agreement with the empirical Tibaldi-Molteni index. The fact that during the blocking event the unstable modes are more localised in the region where the blocking is present is in agreement with results obtained by, e.g., Frederiksen (1997, 2000) using a finite-time normal modes of the tangent linear, and has long been a key element informing the development of operational ensemble weather prediction systems. Very similar conclusions on the spatial structures of the unstable modes can be drawn regarding the properties of the Pacific blockings (Fig. 7b): also in this case, the leading unstable CLVs have larger than usual projection in the Pacific sector when the Pacific blocking is active, with lower than average values before and after the event.

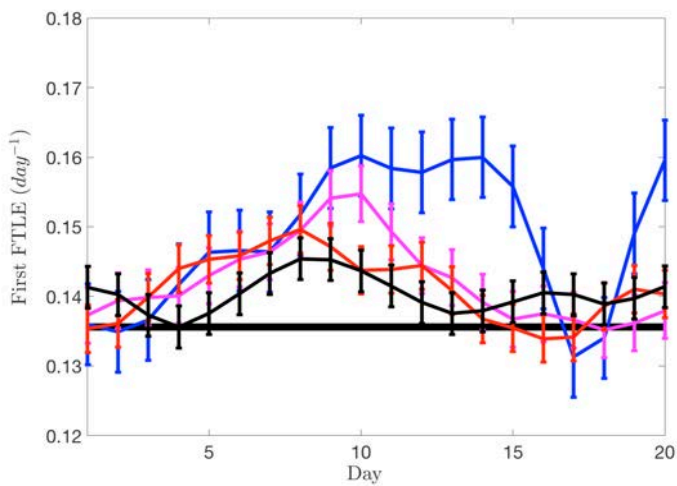

a)

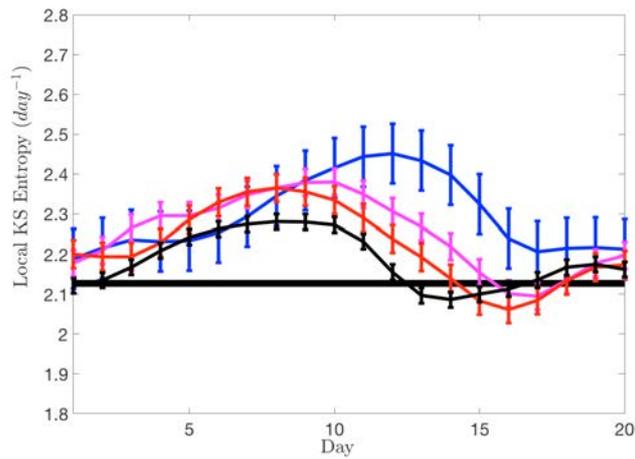

b)

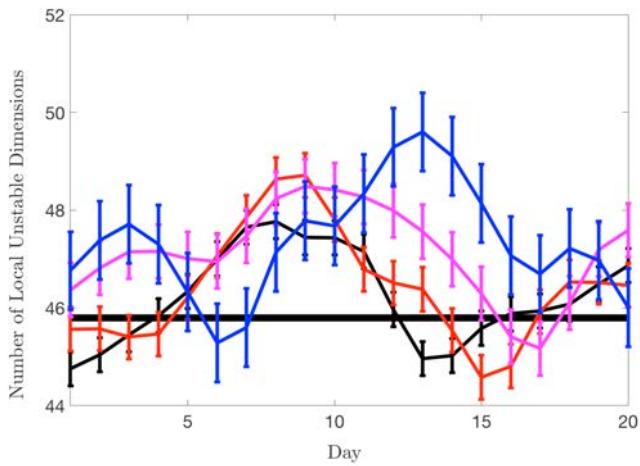

c)

**Figure 8: Same as Fig. 6, but for Pacific blockings.**

We show in Figs. 8a-c for the Pacific blockings the time evolution of the dynamical indicators portrayed in Figs. 6a-c for the Atlantic blockings. In this case, we find about 900 occurrences of 3-day blockings, 280 occurrences of 5-day blockings, 300 occurrences of 6 to 8 day-blockings, and 60 occurrences of 9 to 12-day blockings. Importantly, we find that also in the case of Pacific sector blockings are associated with positive anomalies in all indicators of instability and that longer blockings have higher levels of instability.

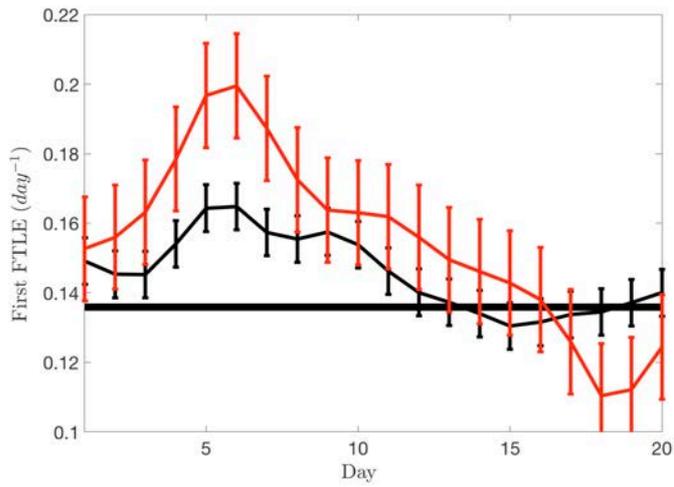

a)

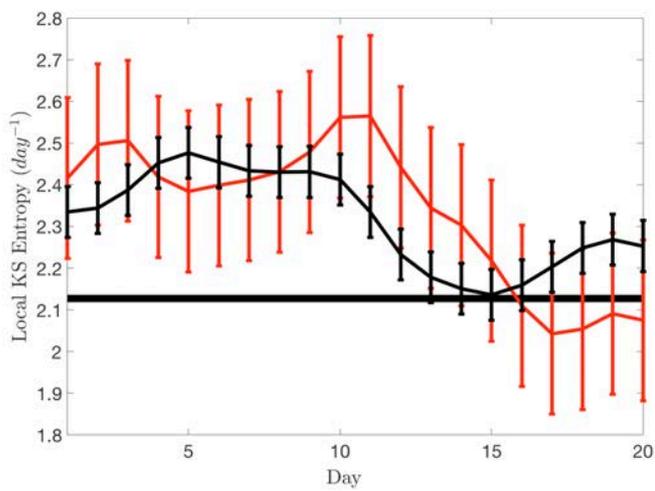

b)

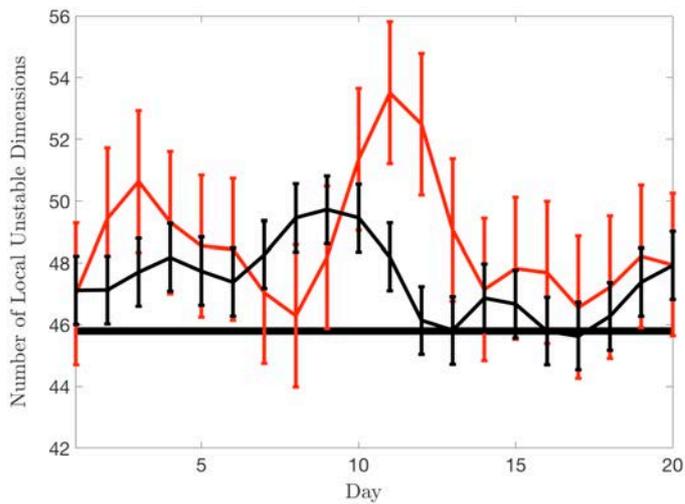

c)

**Figure 9:** Same as Fig. 6, but for global blockings.; results are portrayed only for 3-day and 5-day blocking events.

But, indeed, the life cycle of Pacific blockings is different than in the Atlantic case, and does not conform to what proposed before on the prevalence of instability at the onset and decay of the events. One finds that the onset of the Pacific events coincides approximately with the moment when the instability becomes stronger than the long-term average value. The instability then peaks in the second half of the life of the event, before rapidly decreasing and becoming smaller than the long-term average when the event ends. Qualitatively similar behaviour is indeed found when looking at all the dynamical indicators used here.

Note that, despite the fact that we are considering a severely simplified model of the atmosphere, one can interpret the differences in the life cycle of Pacific vs. Atlantic blockings to the results by Nakamura et al. (1997), where it is argued that Atlantic blockings are mainly forced and modulated by low-frequency patterns, while high-frequency, synoptic forcings are responsible for the formation of Pacific blockings. One can then interpret the Pacific blocking as a resonant response to a forcing, which decays after the forcing has reached its peak.

Finally, we look into the properties of global blockings., which are rather special and rare. Results are shown in Figs. 9a-c. While the statistics of these events is admittedly weaker than for the case of regular Pacific and Atlantic blockings, we can still draw useful conclusions. We report results on 3-day blockings (90 events) and 5-day blockings (9 events). These events typically have a higher degree of instability than blockings of corresponding length occurring in either sector, and, indeed, the longer-lived blocking, the higher their instability. Looking at their life cycle, they resemble considerably the Atlantic blockings, as the instability peaks at the beginning and at the end of the blocking event; note though that the double peak structure is absent for the 5-day long blockings when looking at the largest FTLE.

We remark that in all cases considered we find that blocked states feature, on the average, conditions of anomalously high instability, and that such an anomaly is larger when we consider longer-lived blockings. This indicates that persistent blocking conditions are associated with specific, very unstable regions of the phase space. We will clarify this point by taking the angle of UPOs in the next section.

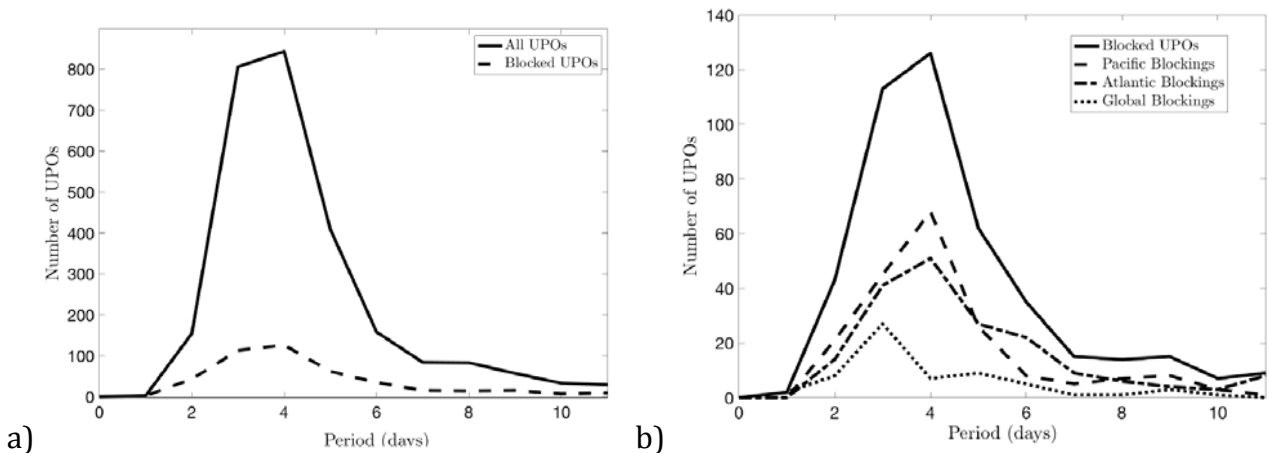

**Figure 10: Number of detected UPOs vs their prime period. a) All UPOs and UPOs featuring blocked states. b) Detail of the UPOs with blocked states.**

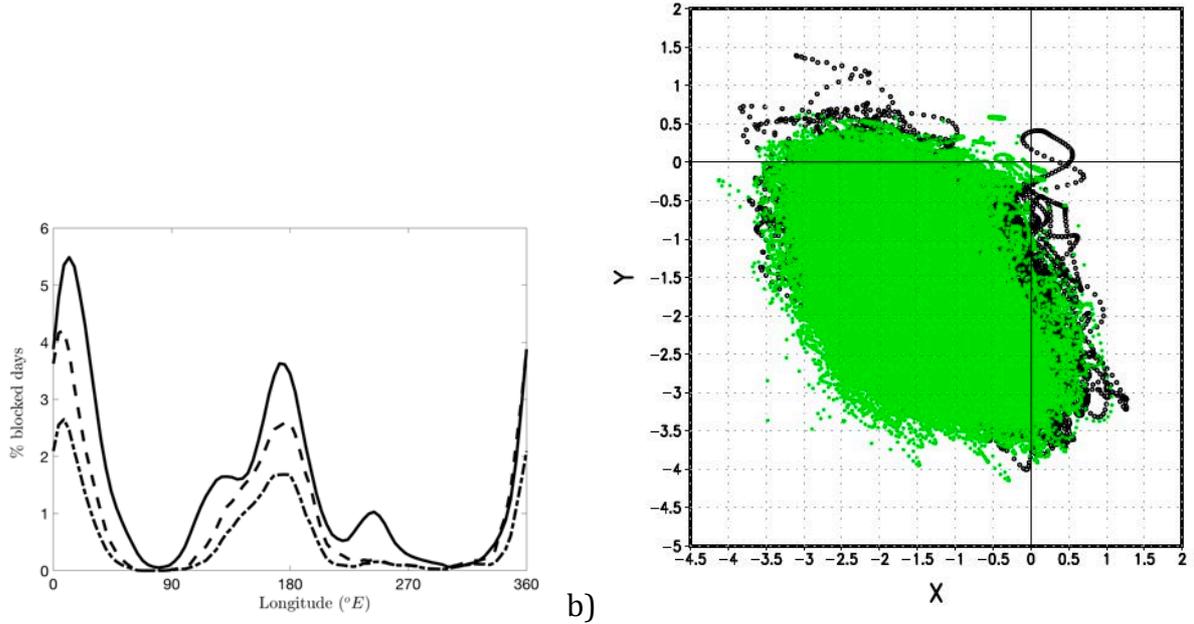

**Figure 11: a)** Representation of the geographical pattern of blocking events by UPOs. Solid line: a) Same as Fig. 3a; dashed line: statistics collected from all detected UPOs, using equal weighting; dash-dotted line: statistics collected from all detected UPOs, using the weighting described in the Appendix. **B)** Projection of all detected UPOs (green) and of the model trajectory (black) on the plane of Tibaldi-Molteni indices $GHSN(\lambda, t)$ (x-axis) and $GHGN(\lambda, t)$ (y-axis) estimated at $\lambda = 0, \delta = 0$ and normalized by $12\text{m}/^0\text{lat}$. The Atlantic blocking conditions for $\lambda = 0, \delta = 0$ correspond to the region (x>0, y<-1).

### 3.3 Unstable Periodic Orbits, Atmospheric Modes, and Blockings

We now look at blockings as modes of atmospheric variability using the mathematical technique of UPOs, because they provide a modal decomposition of the dynamics even in a turbulent regime. As discussed in the Appendix, the theory of dynamical systems indicates that UPOs populate densely the attractor chaotic systems. As a result, a trajectory can be described as being repelled between neighbourhoods of different UPOs, so that locally the dynamical properties of the trajectories can be identified with those of a neighbouring UPO. One can see UPOs as the dynamical, time-dependent equivalent of weather analogues. Our goal here is not to use UPOs to make quantitative statements, but rather to better understand the model properties at a qualitative yet rigorous level.

In Fig. 10a we report the statistics of prime periods of the UPOs we have been able to find. We detect a total of 2711 UPOs. As typical - see e.g. Gritsun 2013 - the sample of detected UPOs is strongly biased towards those possessing short periods. Up to a period of about 3 days, the number of UPOs we detect does increase with T. It is extremely challenging as a result of computational complexity of the problem, to find UPOs with a long period. About 15% (in fact, 441) of the detected UPOs are characterised by going through blocked states. Figure 10b shows the detail of the UPOs featuring Atlantic (count of 185), Pacific (count of 192), or global (count of 64) blockings.

It is reasonable to ask whether such a crude sampling of the whole set of UPOs cover in any meaningful way the attractor of the system, as opposed to providing just some *anecdotal* information on the dynamics. A positive answer to this question is supported by Fig. 11a, where we show the statistics of blockings detection longitude by longitude obtained using UPOs that show a blocking event lasting at least 2 days. We portray two curves, one constructed using equal weighting for all UPOs, and one constructed using the weighting valid for the case of Axiom A systems, as discussed in the Appendix. Our goal here is to show the geographical location and the ratio between occurrence of Atlantic and Pacific blockings are qualitatively captured in a reasonable way by our – very limited – set of UPOs. Additionally, Fig. 11b shows that, in the plane spanned by the two variables introduced in Eqs. (5a-b) (we consider $GHGN(\lambda = 0^oE, \delta = 0, t) < -12m/$ °lat and $GHGN(\lambda = 0^oE, \delta = 0, t) > 0$) the set of UPOs we have computed covers, at least qualitatively, the attractor of the system, and is indeed able to represent the occurrence of Atlantic blockings.

Let's now analyse the dynamical properties of all detected UPOs of the system; see Figs. 12a-c. We plot the distribution over all the detected UPOs of the asymptotic (we are following a periodic orbit)– rather than finite time – LEs and related dynamical indicators on each UPO. Indeed, all detected UPOs are unstable and are very diverse in terms of the dimensionality of their unstable manifold. This confirms (Lai et al. 1997, Kostelich et al. 1999, Do and Lai, 2004) that the system studied here features the very serious violation of hyperbolicity associated with the variability of the number of unstable dimensions. If we compare the corresponding panels of Fig. 5 and Fig. 12, we discover that the UPOs we detect are able to explain, at least qualitatively, the heterogeneity of the attractor in terms of all the considered instability indicators. A forward trajectory of the system explores such the dynamical landscape by hopping between the neighbourhoods of very diverse UPOs. This corresponds to the well-known fact in meteorology that predictability depends critically on the state of the system, and that no prediction of the state of weather is complete without predicting as well its future predictability (Palmer, 2000). What shown here suggests that, despite the intrinsic difficulties in sampling adequately the UPOs, they have a great potential for understanding the properties of the atmosphere.

We remark that the two sets of figures should not be compared at face value, but rather in a qualitative sense, because Figs. 5a-c show the statistics averaged over the attractor of finite-time quantities, while Figs. 12a-c show the statistics according of the detected UPOs, i.e. with no use of appropriate weighting, of asymptotic dynamical quantities.

We now want to investigate to what extent blocking events are associated with specific modes of the circulation. In Figs. 13a-c we compare the statistical properties of UPOs corresponding to typical, zonal patterns to those featuring blockings. What follows applies for both Pacific and Atlantic blockings. Orbits including short-lived blocking events (duration equal or less than two days) are barely distinguishable from the statistics of all UPOs. When looking at UPOs featuring blockings whose lifetime is equal to or longer than three days, we find confirmation that blocked states are anomalously unstable, and that the lifetime of a blocking event correlates positively with its average instability. The estimates of their KS entropy are biased substantially high compared to the statistics of all UPOs. The special nature of instability during blocking events is better understood when looking at the properties of UPOs that are in perennial blocked state. For these UPOs the mean and the standard deviation of the first FTLE, of the local values of the KS entropy, of the KY dimension, and of the number of unstable dimensions are much higher than for the other UPOs.

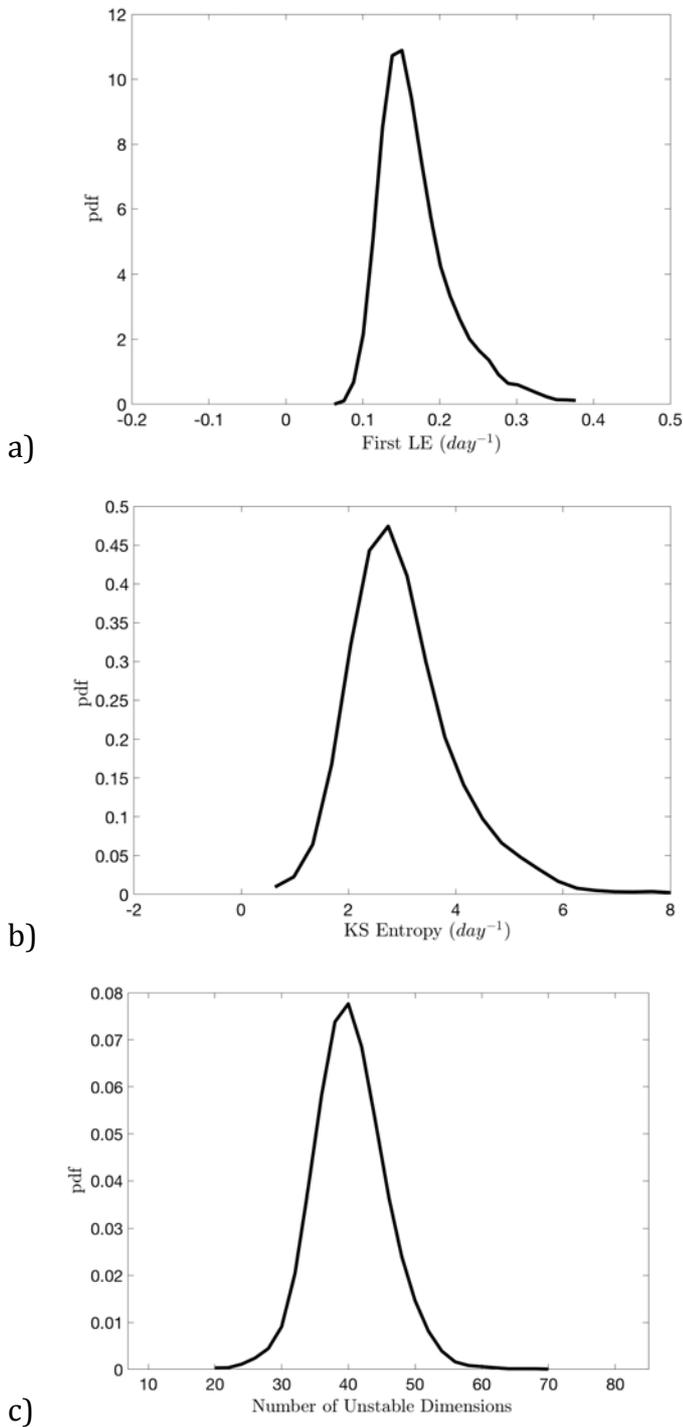

a)

b)

c)                                                                                               d)

**Figure12: Statistics of all detected UPOs: first finite-time Lyapunov exponent (Panel a); local estimate of the Kolmogorov-Sinai entropy (panel b); number of local unstable dimensions (Panel c).**

The onset of blocking event takes places when the trajectory of the system hops from the neighbourhood of a typical (associated with zonal flow) UPOs to the neighbourhood of one of these special, perennially blocked UPOs, which correspond to very special and very rarely visited modes of the system. Longer blockings result come unusually long-lasting periods in which the orbits are very close to such modes. The blocking event ends when, the trajectory hops away from neighbourhood of these special UPOs. Since these modes are very unstable, the time a typical trajectory spends near them is, by definition, low. High degree of instability and low recurrence are intimately related, as a result of the properties of the UPOs; see the Appendix for details. The case of global blocking is not portrayed here in any figure because the number of associated UPOs is low, so that it is hard to extract any information that is meaningful in a statistical sense.

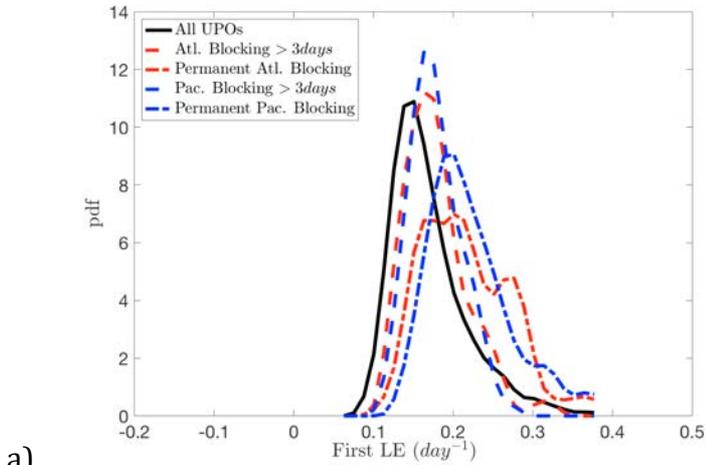

a)

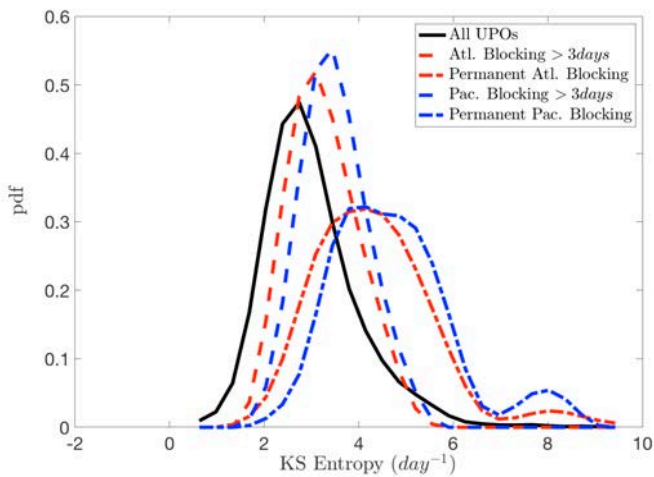

b)

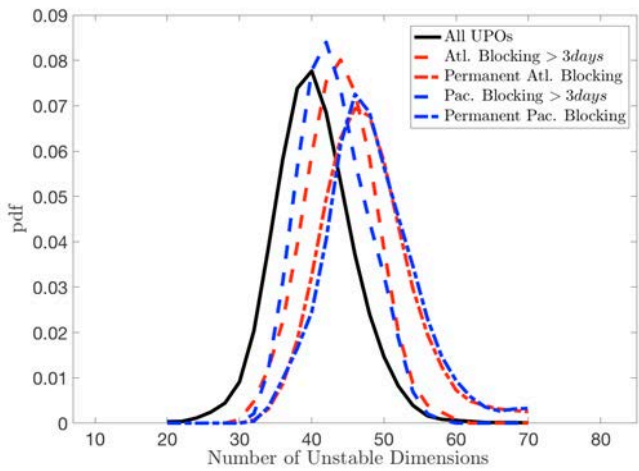

c)                                                                                      d)

**Figure 13: Statistics of UPOs: first finite-time Lyapunov exponent (Panel a); local estimate of the Kolmogorov-Sinai entropy (panel b); number of local unstable dimensions (Panel c). Black lines: same as Fig. 11a)-c). Red dashed lines: UPO with Atlantic blocking patterns with duration longer than 3 days. Red dash-dotted line: UPO with perpetual Atlantic blocking. Blue dashed lines: UPO with Pacific blocking patterns with duration longer than 3 days. Blue dash-dotted line: UPO with perpetual Pacific blocking.**

## 3.4 Teleconnection Patterns, Instability, and Blockings

Previous investigations (Vannitsem 2001) found that very unstable conditions are associated with specific atmospheric patterns, which are reminiscent of the negative phase of the PNA. By computing conditional averages, we find here – see Fig. 14a-b - that the principal component (PC) describing the phase of EOF1 (which strongly project on the negative PNA, see Fig. 2c) is positively correlated with the number of local unstable directions. This indicates that the reduction in the intensity of the jet and the enhancement of its wavy pattern correspond to a stronger instability of the atmosphere as a whole. Other authors (Croci-Maspoli et al. 2007) proposed that negative phases for PNA and/or NAO favor the presence of blockings. We find here that when blocked conditions – Atlantic, Pacific, or global - are present, the EOF1 is preferentially in the positive phase. This provides a strong link between anomalous instability of the atmosphere, presence of blockings, and conditioning given by the PNA teleconnection patterns.

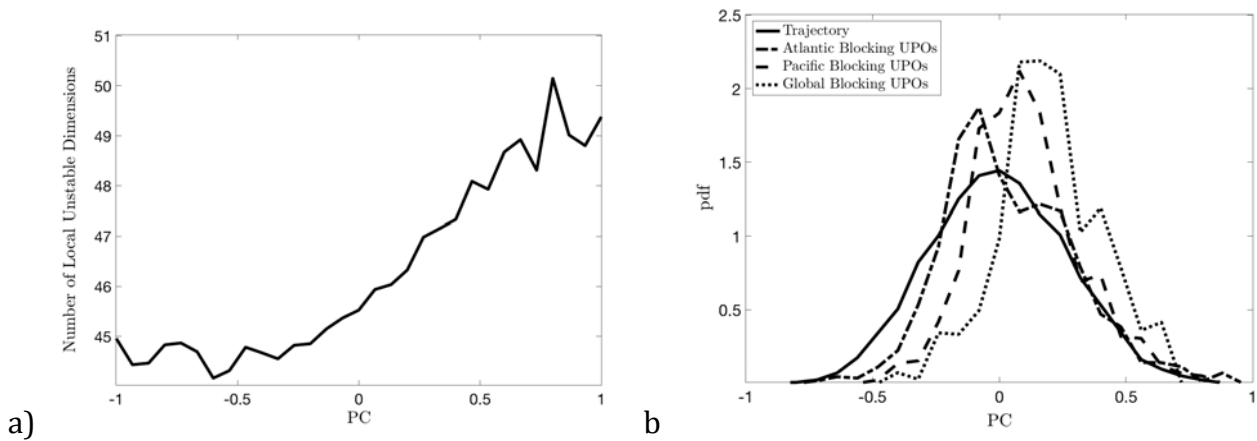

a)                                                                                    b)

Figure 14: a) Number of local unstable directions as a function of the phase of the EOF1 (show in Fig. 3c). b) Pdf of the phase of EOF 1stratified according to the state of the system. Black line: Full trajectory. Dash-dotted/dashed/dotted line: Conditions of Atlantic/Pacific/global blocking, respectively.

## 4. Conclusions

Blocking events provide one of the most relevant and most studied example of weather patterns associated with a large portion of the low-frequency variability of the atmosphere, and sometimes lead to dangerous and high-impact events that affect human and environmental welfare. Despite many years of continuous progresses, numerical weather prediction systems have a comparatively low (yet improving) skill in predicting the onset and decay of blockings, and state-of-the-art climate models have a comparatively hard time in providing a statistics of blockings – in terms of temporal prevalence and geographical location – that fits well with observations. The understanding of how climate change will impact the statistics and dynamics of blocking events is far from being settled. There is a vast and extremely meaningful body of literature dedicated to understanding the physical and meteorological processes responsible for the onset, persistence, and decay of blockings events, using theory, models of various degrees of complexity and observational data. Clearly, there is no simple recipe behind blocking events, and it is hard to have a comprehensive picture of this phenomenon, able to account also for the differences one finds when looking at different geographical locales (Tibaldi and Molten 2018).

In this paper, we have tried to propose a new mathematical framework for understanding the structural properties of blocking events, taking inspiration from the classical low-order simplified models, and using the machinery of modern ideas and methods of dynamical systems theory and statistical mechanics. We have focused on three aspects:
a) how well they can be predicted and how they influence the predictability of the atmosphere;
b) whether they can be associated with *modes* of the atmosphere;
c) why numerical modelling seems to be not so successful in simulating blocking events.

Our investigation has been performed the relatively simple Marshall-Molteni (1993) model of the atmosphere. This model has a fairly good representation of the dynamics of the mid-latitudes and has been widely used for studying the synoptic and low frequency variability. In this work, we have used a low-resolution and hemispheric version of the model because the computational cost incurred in evaluating the mathematical objects used for addressing the questions above. While, clearly, many aspects of the real world are missing from our modelling tool, we maintain that our findings are robust and should be explored using more complex models. We remark that we are proposing a new angle on the problem, and we definitely do not expect to provide a comprehensive answer on the properties of such a complex phenomenon as blocking (Masato et al. 2012), whose phenomenology and aetiology have many more facets than what we have been able to explore here.

Using the formalism of FTLEs, we confirm and substantially extend the findings obtained by Schubert and Lucarini (2016). The first robust result is that, indeed, blocking conditions are associated with higher instability than typical conditions, no matter whether we look at Atlantic, Pacific, or global blockings. Global blockings are very rare and characterised by a very large instability, even compared to the other two sectorial blockings. A second robust result is that the longer is the lifetime of the blocking event, the stronger is its instability.

When looking at the life cycle of the blockings, differences emerge between the two sectorial blockings. In the case of Atlantic blockings, predictability is, on the average, lowest at the onset and decay of the blocking, with a local maximum in the mature phase. Dynamical indicators such as the number of unstable dimensions, the size of the first FTLE, and the local estimate of the KS entropy peak at the beginning and at the end of the blocking events, and dip in the mature phase, when predictability is slightly enhanced. This result is in agreement with a dynamical scenario of formation and then decay of a pattern. In the case of Pacific blockings, predictability is typically at a minimum in the mature stage of the blocking, while instability is lower at the beginning and at the end of each event. In this case, one could interpret the onset and decay of blocking as resulting from the competing effect of external forcings and mechanisms of dissipation. We remark here that the analysis of the tangent space provides only limited information on the predictability (associated with infinitesimal perturbations only) compared to what is deemed useful in the weather forecasting practise, where specific measures of skills able to account for error growth well beyond the linear regime need to be considered.

Faranda et al. (2016, 2017), through a brilliant use of extreme value theory, identified blockings as regimes with higher instability, as defined by a higher local dimension of the atmospheric attractor than usual conditions. They also proposed that, in a reduced order representation of the mid-latitude atmosphere, the blocked state could be seen as a repelling fixed point, contrary to classical investigations performed using low-order models (see e.g. Charney and DeVore 1979; Benzi et al. 1986, Ghil 1987; Mo and Ghil 1988; Benzi and Speranza 1989).

The mathematical machinery of UPOs allows putting the idea of regimes in more solid mathematical grounds, because the decomposition of the invariant measure of the dynamical systems describing the evolution of the atmosphere obtained though the evaluation of its UPOs allows one to rigorously define the true, nonlinear modes of variability. Cvitanovic (2013) suggests UPOs can be instrumental for reformulating fluid dynamical turbulence and looking at it through a different lens. We propose here that this applies as well for the case of atmospheric, ocean, and climate dynamics.

We discover that Atlantic, Pacific, and global blockings are indeed associated with a special class of UPOs, which have a higher of instability with respect to those describing zonal configurations of the atmospheric flow. This agrees with the findings obtained looking at the tangent space along the trajectory and clarifies that a blocking event occurs when the trajectory enters neighbourhood of one or more UPOs associated with blockings and persists there. If a trajectory persists for a long time near such a bundle of UPOs, it will pick up a very high degree of instability. We confirm here the presence of a strong link between anomalous instability of the atmosphere, presence of blockings, and conditioning given by a PNA-like teleconnection patterns.

Since these UPOs are anomalously unstable (and are numerically a minority), the permanence of the trajectory in their vicinity is rather short and, indeed, the atmosphere is only relatively rarely in a blocked state. The theory of UPOs relates the presence of a high degree of instability for an UPO with to the low probability of an orbit of being in its vicinity: in some sense, blockings are relatively rare *because* they have higher instability than typical flow configurations. Global blockings are extremely rare, as they much more unstable than sectorial blockings or zonal flow configurations.

Therefore, we can interpret the transitions between different weather regimes as the system jumping between different bundles of UPOs. The finite-state Markov chain modelling approach for the study of weather systems (Ghil 1987; Mo and Ghil 1988; Vautard et al. 1990; Kimoto and Ghil 1993a,b) can be seen as a severely coarse-grained view of what described here, *microscopically*, at the level of UPOs. Note that the procedure of coarse-graining is responsible for the loss of markovianity (Franzke et al. 2008, 2011; Tantet et al. 2015).

We remark that the laboratory experiments performed using a rotating annulus by Weeks et al. (1997) and Tian et al. (2001) showed that the dynamical regime reminiscent of atmospheric blocking did feature, for a nontrivial parametric range, low-frequency modulations. These modulations might result from the kind of special UPOs associated with blockings that we have found here, albeit in a less turbulent regime. Our findings are also in close correspondence with what presented by Ghil et al. (2002), who described low frequency variability as closely associated with the so-called *ghost limit cycles*, unstable remnants of limit cycles still detectable in more turbulent regime than the one where the Hopf bifurcation takes place.

In future investigations, following Schubert and Lucarini (2015, 2016), we aim at analysing the energetics á la Lorenz (1967) of the blockings UPOs in order to assess the relative role of barotropic and baroclinic conversion in defining the instability of these modes and to understand what differs with respect to the usual non-blocked conditions, trying to reconcile the Charney and DeVore (1979) and Charney and Straus (1980) pictures of low frequency variability, and along the lines of Frederiksen and Bell (1990). We also wish to study the transitions between weather regimes in terms of hopping between different bundles of UPOs, building upon the results obtained with a different approach by, e.g., Oortwin and Barkmeijer (1995), Oortwin (1998), and Jiang et al. (2011).

Clearly, we also wish to test the sensitivity of our results to changes in the resolution of the Marshall-Molteni model. We will also look in detail at whether blockings UPOs are characterised by the existence of a possibly approximate functional relationship between streamfunction and QG potential vorticity, as envisioned by the modons' theory (Butchart et al. 1989). We also expect that the implementation of more sophisticated methods for blocking detection would be helpful in improving the quality of our findings.

UPOs differ widely in terms of the dimension of their unstable manifold, which explains the substantial heterogeneity of the instability properties of the tangent space. Physically, this means that in the attractor there are regions that are very different in terms of available energy for conversion via the baroclinic and barotropic channels, which explains why predictability varies so much in the atmosphere. At mathematical level, the variability of the number unstable dimensions is a serious violation of hyperbolicity (Lai et al. 1997, Kostelich et al. 1999, Do and Lai, 2004). This has major implications in terms of our fundamental ability to numerically simulate the atmospheric dynamics, because, numerically simulated trajectories do not typically shadow for long time the true ones. This seems to be a structural issue dealing with numerical modelling and prediction, which comes on top of the well-known issue of chaos, and which might affect our ability to have a high predictability of the first and of the second kind in the sense of Lorenz.

As far as we are aware, such a fundamental lack of structural stability had never been discussed in the context of geophysical flows. We maintain that this property is not model-specific, but rather a generic and robust feature of weather and climate. Given the extremely large variability of the number of unstable dimensions for UPOs associated with blocking events, the prediction of blocking – and of the response of their statistical properties to changes in the system's parameters – might be indeed affected by such a basic mathematical issue. Additionally, an educated guess is that the lack of structural stability makes the statistics of blockings produced by climate models extremely sensitive to the specific choice of deterministic and stochastic parametrizations (Berner et al. 2017) used for representing the impact of small, unresolved scales on the resolved ones. Indeed, Kondrashov et al. (2006) showed that the coarse grained version of the Marshall and Molteni (1993) model featured structural changes in the statistics of weather patterns (from uni- to multimodal) as the parameter controlling the intensity of the stochastic parametrization was altered. The lack of structural stability might partly explain why the statistics of blockings undergoes strong modulations on interannual and multidecadal scales as a result of relatively weak forcings (Häkkinen et al. 2011; Rimbu and Lohmann 2011; Rimbu et al. 2014).

We suggest that the presence of a (strong) variability of the number of unstable dimensions might have impacts on the efficiency of the recently proposed strategy of data assimilation restricted to the unstable and neutral space, defined by the CLVs featuring non-negative LEs (see e.g. Trevisan and Uboldi 2004; Trevisan et al. 2010; Bocquet and Carrassi 2017). Indeed, the need of performing assimilation in a state space that includes also some stable modes has been recently discussed in Grudzien et al. (2018). We believe that the reason for this is that, in some regions of the phase space, some (or even many) dimensions of the stable space (defined by the CLVs featuring negative LEs) might feature a (finite-time) positive growth rate for the perturbations initialised on the corresponding CLVs. Such finite-time, yet possibly locally important instabilities might then be neglected by assimilation procedures that considers only on the unstable space.

Very weak perturbations applied to a uniformly hyperbolic system leads to a small change in its measure (Ruelle 2009), associated with small deformations to the structure of the UPOs. In absence of structural stability, even small perturbations can lead to drastic changes in the UPOs (some UPOs are generated and others disappear), associated with complex set of bifurcations. We propose that the difficulty in predicting the response of blocking events to climate change might be linked at a very fundamental level, apart from the many physical complexities of the real climate that we cannot describe in this model, to the lack of structural stability. Note that this lack of robustness is not in contrast with the possibility that response theory might predict well the climate response to forcings, if one considers sufficiently coarse-grained quantities (Ragone et al. 2016, Lucarini et al. 2017).

The understanding of such a lack of robustness and of the dynamical complexity in the atmosphere, as well as of the mathematical nature of blocking events is worth exploring by taking the point of view of time-dependent and random dynamical systems - see, e.g. Chekroun et al. 2011, Ghil 2017), and by extending the approach presented here to numerical models of the atmosphere either able to describe dynamics on a broader range of scales – e.g. using a primitive equations dynamical core – or incorporating a larger variety of physical processes – e.g. through, even minimal, parametrizations. It is also worth expanding this analysis in the direction of coupled atmosphere-ocean models, in order to be able to decompose the dynamics of climate in its nonlinear modes of variability. We might be able to associate specific UPOs to relevant coupled oceanic-atmospheric modes, and have a different angle for understanding their response to climate change. Indeed, the flexible and customizable atmospheric model PUMA (Frisius et al. 1998) and coupled atmosphere-ocean model MAOOAM (De Cruz et al. 2016), are, respectively, excellent candidates as tools for pursuing such research lines. We remark that performing these investigations will be exciting and well as challenging, as it will require using more efficient algorithms and taking advantage of better computing resources than done in the preliminary work presented here.

**Acknowledgments**

The authors acknowledge the support provided by the Royal Society (UK)– RFBR (Russia) Bilateral grant (Royal Society Grant: IEC\R2\170001; RFBR project 17-55-10012. VL acknowledges the support provided by the Horizon2020 projects Blue-Action (grant No. 727852) and CRESCENDO (grant No. 641816). The authors have benefitted from scientific exchanges with P. Cvitanovic, J. Dorrington, V. Dymnikov, M. Santos, T. Woolings, and J. Yorke, and from the constructive criticism by one anonymous reviewer and by M. Ghil. This paper has been partly prepared during the Advanced Workshop on Nonequilibrium Systems in Physics, Geosciences, and Life Sciences held at ICTP, Trieste, on May 15-24 2018.

## Appendix: Mathematical Background

We give here a rather informal introduction to some mathematical background that is essential for the understanding of the paper. The reader who has solid knowledge of dynamical systems theory is encouraged to skim through or skip entirely this appendix.

### Dynamical Systems and Their Invariant Measure

Let's consider a smooth autonomous chaotic continuous-time dynamical system acting on a smooth compact manifold $\mathcal{M}$ of dimension N evolving from an initial condition $x_0$ at time $t = 0$. We define $x(t, x_0) = \Pi^t(x_0)$ its state at a generic time $t$, where $\Pi^t$ is the of evolution operator. The evolution operator obeys the semigroup property, so that $\Pi^\tau = \Pi^{\tau-s}\Pi^s \ \forall s \in \mathbb{R}_0^+$. We also define $O(x(t, x_0)) = O(\Pi^t(x_0)) = S^t(O(x_0))$ the Koopman operator describing the evolution of a general observable $O(x)$ after a time $t$. The Koopman operator inherits the semigroup properties in a natural way. The corresponding set of differential equations can be customarily written as

$$\frac{dx(t)}{dt} = F(x(t)) \qquad (A1)$$

where $F(x) = d\Pi^s(x)/ds$. Let us define $\Omega \subset \mathcal{M}$ as the compact attracting invariant set of the dynamical system. We assume that we can define the associated ergodic physical (Sinai-Ruelle-Bowen) measure $\nu$ (Eckmann and Ruelle 1985), with support $\Omega = supp(\nu)$. We define the expectation value of an observable $\Phi$ as follows:

$$\nu(\Phi) = \langle \Phi \rangle_0 = \int \nu(dx)\Phi(x) = \lim_{t \to \infty} \frac{1}{t}\int_0^t d\tau \ \Phi(S^\tau x) \qquad (A2)$$

for almost all (in the Lebesgue sense) initial conditions $x$ belonging to the basin of attraction of $\Omega$, where in the last equality we have used the property of ergodicity.

### Lyapunov Exponents

Let us introduce the characteristic exponents describing the asymptotic behaviour of infinitesimal perturbations from a background trajectory. See a comprehensive treatment in Eckmann and Ruelle (1985) and Ruelle (1989). Let $J_t(x) = \nabla_x F(\Pi^t x)$ be Jacobian matrix of the flow at time $t$ with initial condition $x \in \Omega$. We define the matrix $L_t(x) = J_t^T(x)J_t(x)$. The Osedelet (1968) theorem ensures us that the matrix

$$\Lambda(x) = \lim_{t \to \infty} (J_t^T(x)J_t(x))^{\frac{1}{2t}} \quad (A3)$$

exists and that its eigenvalues $\Lambda_i(x)$, i = 1,2 ..., N are constant almost everywhere (with respect to $\nu$), so that the $x$-dependence can be dropped. We define as $\lambda_i = \log \Lambda_i$, i = 1,2 ..., N the spectrum of Lyapunov Exponents (LEs) of the system. Customarily, they are ordered by size - $\lambda_1 \geq \lambda_2 \geq \cdots \geq \lambda_N$, − and for a chaotic system $\lambda_1 > 0$[1]. When considering a flow, we have that at least one (and only one in the case of nonuniform hyperbolic systems, see Katok and Hassenblatt 2003) of the Lyapunov exponents vanishes because it corresponds to the direction of the flow. If n defines the index of the smallest positive LE, we say that the dimensionality of the unstable manifold is $n$. Note that $\sum_{i=1}^{N} \lambda_i = \int \nu(dx) \nabla_x \cdot F(x)$, i.e. the sum of the LEs is equal to the expectation value of the divergence of the flow. While the LEs are asymptotic quantities, one can also consider the finite-time LEs (FTLEs) $\lambda_1(x,t) \geq \lambda_2(x,t) \geq \cdots \geq \lambda_N(x,t)$, which are the eigenvalues of $\Lambda(x,t) = (J_t^T(x)J_t(x))^{1/2t}$ and depend explicitly on $x$ and $t$. These are referred to as backward FTLEs. Clearly one has that $\lambda_j = \lim_{t \to \infty} \lambda_j(x,t)$ for x $\nu$-almost everywhere. Note that the ensemble (or long-time average along the trajectory) of each $\lambda_j(x,t)$ gives $\lambda_j$.

Chaotic systems describing nonequilibrum, forced and dissipative systems feature a negative sum of their LE. Therefore, the set $\Omega$ has zero N-dimensional Lebesgue measure. Instead, one can introduce generalized notions of (fractal) dimension in order to provide quantitative characterizations of $\Omega$. While the theory of Renyi dimensions gives an overarching method to study the properties of $\Omega$, the Kaplan-Yorke conjecture proposes a definition of the fractal dimension of $\Omega$ as follows:

$$D_{KY} = m + \frac{\sum_{i=1}^{m} \lambda_i}{|\lambda_{m+1}|} \leq N \quad (A4)$$

where $m$ is the largest number such that $\sum_{i=1}^{m} \lambda_i \geq 0$. The LEs can be used to find an explicit expression for the KS entropy of the flow, which provides the rate of creation of information due to the system's sensitive dependence on initial conditions; the Pesin theorem says that

$$h_{KS} = \sum_{\lambda_i > 0} \lambda_i, \quad (A5)$$

which indicates that $h_{KS}$ coincides with the rate of volume expansion along the unstable dimensions of the flow; for chaotic systems one has that $h_{KS} > 0$. . We construct here a local version - $h_{KS}(x,t)$ of- $h_{KS}$ by taking the sum of all backward FTLEs $\lambda_j(x,t)$ whose corresponding LEs $\lambda_j$ are positive. The ensemble (or long-time average along the trajectory) of $h_{KS}(x,t)$ gives $h_{KS}$.

---

[1] See an additional condition for chaos below when discussing Unstable Periodic Orbits.

**Covariant Lyapunov Vectors**

The Covariant Lyapunov Vectors (CLVs) provide a covariant basis $\{c_1(t), c_2(t), ..., c_n(t)\}$ describing the solutions to the following system of linear ordinary differential equations:

$$\dot{y} = J_t(x)y \quad (A6)$$

where $J_t(x)$ has been defined above. The main property of the basis of CLVs is that setting $c_j(t_1)$ as initial condition for $y$ at time $t_1$ in the evolution equation of the tangent space, at time $t_2 > t_1$ the solution is parallel to $c_j(t_2)$ and in the limit for $t_2 \to \infty$ the average growth (or decay) rate of its amplitude is given by the $j^{th}$ LE $\lambda_j$. The finite-time growth rate of the $j^{th}$ CLV over a time scale $t$ along the orbit with initial position $x$ defines the covariant FTLE $l_j(x,t)$, where $\lambda_j = \lim_{t \to \infty} l_j(x,t)$ for x $\nu$-almost everywhere.. Note that, by construction, $\lambda_1(x,t) = l_1(x,t)$. See discussion in Kuptsov and Kuznetsov (2018) for a detailed discussion of the conceptual differences between the covariant - $l_j(x,t)$ - and backward - $\lambda_j(x,t)$ – FTLEs.

The CLVs corresponding to positive (negative) LEs span the unstable (stable) tangent space, while the CLV corresponding to the vanishing LE (assuming it is unique) is oriented along the direction of the flow and spans the neutral direction of the tangent space. Efficient algorithms for identifying the CLVs were first determined independently by Ginelli et al. (2007) and by Wolfe and Samuelson (2007). See a comprehensive review in Froyland et al. (2013). Hyperbolic flows are such that the stable and the unstable tangent spaces have no tangencies, so that all trajectories on the attractor are saddle. Hyperbolic flows are structurally stable (Katok and Hasselblatt 2003), possess a physical measure, and are physically relevant, as clarified by the chaotic hypothesis (Gallavotti 2014).

By counting the number of covariant FTLEs we define the local (in space and time) number of unstable dimensions $n(x,t)$. Such quantity is computed at temporal resolution of one time step, and daily values are constructed as averages over 40 consecutive time steps. In this case the ensemble average of the finite-time estimators does not coincide with the asymptotic value, because of lack of linearity in the definition of the number of unstable dimensions

**Unstable Periodic Orbits**

A periodic orbit of period $T$ is defined as follows:

$$\Pi^T(\vec{x}) = \vec{x} \quad (A7)$$

where, as explained below, such a representation is not unique. First, if the previous equation is verified, then we also have $\Pi^{nT}(\vec{x}) = \vec{x}, \forall n \in \mathbb{N}$, so that from now on when we talk about the period of an orbit we implicitly refer to its prime period $T$ (unless otherwise stated). Secondly, by the semigroup property, we have that $\Pi^T(\vec{y}) = \vec{y}$ if $\vec{y} = \Pi^s \vec{x}$, for any choice of $s$.

The attractor of a chaotic system is densely populated by unstable periodic orbits (UPOs). UPOs provide the so-called skeletal dynamics. One can think the chaotic trajectory to be always near at least one UPO, but never following any of them indefinitely, because of their instability. This implies that periodic orbits can approximate any trajectory in the system with an arbitrary accuracy, and all statistical characteristics of the system can be calculated from the full set of periodic orbits (Auerbach et al. 1987; Cvitanovic 1988, 1991; Cvitanovic et al. 2016). Therefore, through the use of so-called trace formulas, one can formally construct the invariant measure $\nu$ of the system by considering the following expression for the expectation value of any measurable observable $\Phi$:

$$\nu(\Phi) = \lim_{t \to \infty} \frac{\sum_{U^p, p \leq t} w^{U^p} \overline{\Phi^{U^p}}}{\sum_{U^p, p \leq t} w^{U^p}} \qquad (A8)$$

where $U^p$ is a UPOs of prime period $p$, $\overline{\Phi^{U^p}}$ is the average of the observable $\Phi$ taken on the orbit $U^p$, and $w^{U^p}$ is the weight of the UPO $U^p$. In the case of uniformly hyperbolic systems, such a weight, to a first approximation, can be expressed as $w^{U^p} \propto \exp\left(-p h_{ks}^{U^p}\right)$. Therefore, the weight decreases exponentially with the information generated by the system in one period of the UPO. See the derivation in Grebogi et al. (1988) in the case of uniformly hyperbolic discrete maps, the discussion in Cvitanovic (1988), and the comment by Zaks and Goldobin (2010) to Saiki and Yamada (2009) on the importance of using the right weight.

Some investigations suggest that the weighting $w^{U^p}$ above can be used effectively also for more general systems (Lai 1997; Lai et al. 1997), while other authors have proposed the use of heuristic formulas where a different weighting is used (Kazantsev 1998; Zoldi 1998). On the other side, one knows that the number of UPOs of period $t$ grows exponentially with $t$ times the topological entropy (Hasselblatt and Katok, 2003), which provides, roughly speaking, an upper bound to the metric entropy. Therefore, choosing the cut-off maximum period $T_{max}$ at which we truncate the sum in Eq. (A8) is far from being a trivial task. In fact, the long-period UPOs tend to be under-represented in any numerical approximation. In the case of uniformly hyperbolic systems, this seems not to create major problems if one wants to evaluate averages using the formula given in Eq. (A8), while the contribution of long period UPOs might be relevant in more general cases (Cvitanovic 1988). A detailed treatment of the problem can be found in Gritsun (2008) and Cvitanovic et al. (2016). In this paper we do not attempt to evaluate the weights of the detected UPOs, but consider them as building blocks of the system, able to provide a robust qualitative information on its properties.

Equation (A7) has N+1 unknowns (the N coordinates of the initial condition of the orbit and the period of the orbit) and it is in general impossible to solve it explicitly. For a given system, we expect many (in fact, infinite) solutions. We need to resort to numerical methods that update an initial guess of $\vec{x}_0$ and $T_0$ until we obtain $\vec{x}$ and $T$ obeying the equation above. It is useful to provide a brief description of the classical Newton iterative approach, which provides the basis of more advanced search methods. A possible way to choose suitable $\vec{x}_0$ and $T_0$ is to look at a long integration of evolution equation and choose a quasi-recurrence occurring over a period $T_0$, such that $|\Pi^{T_0}(\vec{x}_0) - \vec{x}_0| < \varepsilon$, where $\varepsilon$ is a prescribed value. The iterative procedure then goes as follows. We now write the following equation:

$$\Pi^{T_0+\delta T}(\vec{x}_0 + \delta\vec{x}) = \vec{x}_0 + \delta\vec{x} = \Pi^T(\vec{x}_0) + \partial_T S^T(\vec{x}_0)\delta T + \vec{\nabla}\Pi^T(\vec{x})|_{\vec{x}=\vec{x}_0}\delta\vec{x} \quad (A9)$$

We recall that $J_T(x) = \vec{\nabla}\Pi^T(\vec{x})|_{\vec{x}=\vec{x}_0}$, while, by definition, $\partial_T S^T(\vec{x}_0) = F(\Pi^T(\vec{x}_0))$. We then have:

$$\vec{x}_0 - \Pi^T(\vec{x}_0) = F(\Pi^T(\vec{x}_0))\delta T + (J_T(x)|_{\vec{x}=\vec{x}_0} - 1)\delta\vec{x} \quad (A10)$$

This equation is then supplemented by the condition:

$$\delta\vec{x} \cdot F(\vec{x}_0) = 0 \quad (A11)$$

which says that we update the starting position of the orbit in a linear space orthogonal to the local flow, because the periodic orbit does not change if we move with the flow. Combining Eqs. (A10) and (A11) we can find $\delta\vec{x}$ and $\delta T$. We now define $\vec{x}_1 = \vec{x}_0 + \delta\vec{x}$ and $T_1 = T_0 + \delta T$, and iterate the procedure *hoping* it will converge, meaning that we can define $\vec{y} = \lim_{n\to\infty} \vec{x}_n$ and $\tau = \lim_{n\to\infty} T_n$ such that $\vec{y} = \Pi^\tau(\vec{y})$. Because of the strong nonlinearity, it is often better to use numerically more efficient methods, such as the damped Newton or inexact quasi-Newton method (with line search, multiple shooting, and tensor correction). See Saiki (2007), Crofts and Davidchack (2006) and Cvitanovic et al. (2016) for further inputs.